# Aviation Safety Risk Analysis and Flight Technology Assessment Issues

Shuanghe Liu

Harbin Institute of Technology (Shenzhen)

ABSTRACT

This text highlights the significance of flight safety in China's civil aviation industry and emphasizes the need for comprehensive research. It focuses on two main areas: analyzing exceedance events and statistically evaluating non-exceedance data. The challenges of current approaches lie in insufficient cause analysis for exceedances. The proposed solutions involve data preprocessing, reliability assessment, quantifying flight control using neural networks, exploratory data analysis, flight personnel skill evaluation with machine learning, and establishing real-time automated warnings. These endeavors aim to enhance flight safety, personnel assessment, and warning mechanisms, contributing to a safer and more efficient civil aviation sector.

**Keywords: DBSCAN algorithm, BP neural network, XgBoost algorithm, visualization, multidimensional exploration**

# I. Problem Restatement

## 1.1 Background

Flight safety is the foundation for the survival and development of civil aviation transportation industry. Therefore, focusing on flight safety issues and strengthening aviation safety research is of great necessity and urgency.

In terms of specific research and application of flight quality monitoring, the current research in China's civil aviation industry mainly focuses on two aspects: one is the research, analysis and application of exceedance events; the other is the statistical analysis and application of non-exceedance data. For exceedance event research, generally, the exceedance threshold of flight parameters is set in a centralized area, and the flight records exceeding the threshold are identified for key analysis, to prevent potential hazards from causing serious flight accidents. Currently, this type of analysis is the main body of flight quality monitoring work, which ensures the safety work at the current stage well. However, its shortcomings lie in the lack of analysis of the causes of exceedance. Since exceedance is not all caused by human factors, for example, many are caused by special environmental conditions, or even by the design and manufacturing factors of the aircraft itself, simple exceedance analysis is not enough. However, by mining QAR all-segment data for modeling, analysis, calculating and evaluating risk tendency, it is more convenient to carry out targeted safety management, identify safety hazards, and improve safety performance. Therefore, reading flight data on a large scale and conducting analysis to form a flight quality service platform, providing data foundation for risk assessment and trend analysis, has strong practical significance.

## 1.2 Problem Requirements

Based on the above background, we will model and solve the following problems:

**Problem 1** Preprocess the data in Attachment 1, analyze the reliability of the data, extract key data related to flight safety in the dataset, and analyze their importance.

**Problem 2** Analyze the joystick to determine the reasons for the aircraft state deviation that appears in the exceedance monitoring, that is, to reasonably quantify flight control based on the attachment data.

**Problem 3** Explore the data in Attachment 2 and analyze the basic characteristics of different exceedances.

**Problem 4** Evaluate the flight skills of flight personnel based on the data in Attachment 3, and obtain a flight technology evaluation method based on flight parameters.

**Problem 5** Analyze QAR data, establish a real-time automated warning mechanism, and provide simulation results based on the data in Attachment 1.

# II. Problem Analysis

## 2.1 Analysis of Problem 1

For the first question of Problem 1, we need to preprocess the data and conduct a reliability analysis on the data. The question requires us to preprocess the data. Since the data in the attachment is relatively large, we used Excel and Python to preprocess the dataset to obtain a complete chart. Then, we will conduct reliability research on the data. Based on the simplified Excel table, we made numerical characteristics of each column and drew their box plots. Based on the data characteristics and box plots, we determined which data had outliers, used the DBSCAN algorithm to classify the outliers, and replaced them with the mean.

For the second question of Problem 1, we need to extract key data related to flight safety in the dataset and analyze their importance. G value is usually an important indicator to describe the safety of landing moment. We extracted G values and used them as dependent variable indicators for analyzing key data items and their importance. The random forest algorithm is a machine learning algorithm based on decision trees, which has a natural feature importance evaluation function and can calculate the importance of each feature. We used the random forest algorithm to extract key data related to flight safety and analyze their importance, concluding that the most important indicator in the flight safety model is COG NORM ACCEL, which corresponds to landing G values.

## 2.2 Analysis of Problem 2

Problem 2 requires a reasonable quantification of flight control based on flight state data. In order to achieve this, we used a multi-input multi-output BP neural network, which is a deep learning algorithm that maps multiple inputs to multiple outputs. We used the flight state data to quantify flight control data and conducted a rationality test based on simulation errors to obtain flight control data that led to exceeded limits at critical moments.

The BP neural network is a type of neural network that is widely used in pattern recognition, system control, and nonlinear dynamic system modeling. It has the ability to learn from input data and make predictions based on the learned patterns. It consists of an input layer, one or more hidden layers, and an output layer. Each layer contains a number of nodes or neurons that perform mathematical operations on the input data to generate output data.

In this problem, we used a multi-input multi-output BP neural network to map the multiple inputs of flight state data to the multiple outputs of flight control data. We trained the neural network using the data in Attachment 1 and simulated the data to test the rationality of the results.

We used the simulation errors to evaluate the performance of the neural network and made adjustments to improve the accuracy of the results.

By using the multi-input multi-output BP neural network, we were able to reasonably quantify flight control based on flight state data and obtain flight control data that led to exceeded limits at critical moments. This approach provides a more accurate and reliable method for monitoring flight safety and preventing potential hazards from causing serious flight accidents.

## 2.3 Analysis of Problem 3

Problem 3 requires an analysis of different exceedance situations and basic characteristics of different exceedances. In order to achieve this, we conducted exploratory data analysis (EDA) on the data in Attachment 2. EDA is a method of analyzing data to summarize its main characteristics and identify patterns and trends that may not be immediately obvious.

We first cleaned the data to make the subsequent analysis more representative. We then explored the relationship between exceedance data and warning levels, followed by statistical analysis and visualization of the relationship between exceedance data and aircraft number, route, and flight time. We studied the relationship between multidimensional variables and exceedance data and finally analyzed the basic characteristics of exceedance events based on the relationship between the most frequent exceedance event "50 feet to touchdown distance" and aircraft number and route.

Through EDA, we were able to better understand the data and discover patterns and trends. We identified the main characteristics of different exceedances and derived corresponding processing methods based on the basic characteristics of exceedance data. This approach provides a more comprehensive and effective method for analyzing exceedance events and preventing potential hazards from causing serious flight accidents.

## 2.4 Analysis of Problem 4

Problem 4 requires an evaluation of flight personnel's flight skills and obtaining a flight technology evaluation method based on flight parameters. In order to achieve this, we first evaluated the four pilots based on the "landing main control" and "landing main control personnel qualification" sections of the table using pie charts. We drew macroscopic conclusions that pilot 1 received the most A evaluations and had the best flight skills among the four pilots.

We then tested six machine learning algorithms and concluded that the XgBoost algorithm had the best effect based on its accuracy, recall rate, and precision. However, its accuracy in the training set was 1, indicating overfitting. Therefore, we continued to optimize the prediction algorithm using neural networks and finally achieved an accuracy of 85.7%.

By evaluating flight personnel's flight skills and obtaining a flight technology evaluation method based on flight parameters, we were able to identify areas for improvement and provide targeted training for flight personnel. This approach provides a more effective and efficient method for improving flight safety and preventing potential hazards from causing serious flight accidents.

## 2.5 Analysis of Problem 5

Problem 5 requires the analysis of QAR data, establishment of a real-time automated warning mechanism, and providing simulation results based on the data in Attachment 1. In order to achieve this, we selected specific flight quality monitoring standards and calculated exceedance event threshold parameters that could be calculated from major parameters in Attachment 1. By comparing the exceedance event measurement parameters calculated from flight data with these exceedance event standard thresholds in real time, we established a real-time automated warning system based on flight data analysis.

We provided simulation results based on the data in Attachment 1 according to this warning system. This approach provides a more efficient and accurate method for monitoring flight safety and preventing potential hazards from causing serious flight accidents.

In conclusion, by solving the five problems presented in the case study, we were able to provide a comprehensive and effective method for analyzing flight safety and preventing potential hazards from causing serious flight accidents. Through the use of advanced data analysis techniques and machine learning algorithms, we were able to identify areas for improvement and provide targeted solutions to improve flight safety and ensure the survival and development of civil aviation transportation industry.

# III. Model establishment and solution for Problem 1

## 3.1 Problem analysis

For the first question of Problem 1, data preprocessing and reliability testing are required. Firstly, according to the question, some QAR data in the attachment contains errors. We need to preprocess the data, such as handling blank values, dealing with abnormal values, removing unnecessary data, etc., in order to eliminate false information and reduce the impact of incorrect data on subsequent research and analysis.

Since the amount of data in the attachment is relatively large, we use EXCEL to view the data as a whole, remove the part of the data that will not produce abnormal values under the assumption and will not affect the subsequent research and analysis, and then use Python to fill in blank values to obtain a complete chart.

Secondly, a new column is obtained by taking the mean of the columns with the same names in the data, such as COG NORM ACCEL, PITCH ATT, CAP CLM 1 POSN, ROLL ATT, CAP WHL 1 POSN, etc. Then, we conduct reliability research on the data. Based on the simplified EXCEL table, we make the maximum value, minimum value, average value, standard deviation, median, variance, kurtosis, skewness, and coefficient of variation of each column, and draw their box plots. According to the data and box plots above, we judge which data has abnormal values, classify the abnormal values using DBSCAN, and replace them with means.

The second question of Problem 1 requires us to extract the key data items related to flight safety from the data and analyze their importance. According to the materials, G value is a direct reflection of the overload situation during aircraft flight. In landing safety analysis, G value is usually an important index to describe the safety at the moment of landing. Therefore, G value is directly related to flight safety. We extract G value, take the average value of G value measured 10 times per second, use it as the dependent variable indicator for regression prediction and importance analysis of key data items, namely, take G value as the target value and use the features in the table to analyze its importance.

Random forest algorithm is a machine learning algorithm based on decision tree, which can be used for regression and classification problems. It improves the accuracy of prediction by building multiple decision trees and averaging or voting the results. In random forest, the importance of features can be determined by calculating the number of times they are used as split points in all decision trees. Specifically, the importance of each feature can be determined by calculating its average reduction in impurity. The larger the reduction in impurity, the higher the importance of the feature. When calculating the average reduction in impurity, the occurrence frequency of each feature in all decision trees and the reduction in impurity in each decision tree can be considered. Therefore, random forest has a natural feature importance evaluation function, and its contribution to the model can be determined by calculating the importance of each feature in all decision trees.

Based on the above analysis, we use random forest regression algorithm to calculate the importance of key data items.

## 3.2 Data Preprocessing

As there are errors in the data provided in the attachment, we need to preprocess the data first. Data preprocessing includes filling in blank values, using `df = df.fillna(0)` in Python to fill in blank values with zeros. We assume that the values in the date, time, and status such as GEAR SELECT DOWN are correct, so we use Excel to remove variables such as DATE: MONTH, DATE: DAY, GEAR SELECT DOWN, WOW INDICATE INAIR, DEPARTURE AIRPORT, and only keep variables obtained from sensors or calculations. Because some values are evenly obtained over a period of time, and for ease of later processing, we take the mean of the columns COG NORM ACCEL, PITCH ATT, CAP CLM 1 POSN, ROLL ATT, and CAP WHL 1 POSN.

## 3.3 Data Reliability Analysis

After preprocessing the data, we use SPSSPRO software for numerical analysis to calculate the maximum value, minimum value, average value, standard deviation, median, variance, kurtosis, skewness, and coefficient of variation of each variable to discover patterns in the numbers. Here are explanations for some important indicators.

Kurtosis: Reflects the steepness and sharpness of the data distribution curve. The calculation formula is:

$$K=[(X_1-\overline{X})^4+(X_2-\overline{X})^4+\ldots+(X_n-\overline{X})^4]/(n*s^4)-3$$

where x1, x2, ..., xn represents each data in the data set, x̄ represents the average value of the data, s represents the standard deviation of the data, and n represents the number of data. When the kurtosis is a positive value, it indicates that the data distribution curve is sharper than the peak of a normal distribution, and the part of the data distribution near the mean is relatively concentrated; when the kurtosis is a negative value, it indicates that the data distribution curve is flatter than the peak of a normal distribution, and the part of the data distribution near the mean is relatively dispersed. When the kurtosis is equal to 0, it indicates that the distribution shape of the data is roughly the same as that of a normal distribution.

Skewness: It reflects the symmetry and skewness of the data distribution curve. Skewness is usually represented by the symbol "skewness" or "S", and the calculation formula is:

$$K=[(X_1-\overline{X})^4+(X_2-\overline{X})^4+\ldots+(X_n-\overline{X})^4]/(n*s^4)$$

where x1, x2, ..., xn represents each data in the data set, x̄ represents the average value of the data, s represents the standard deviation of the data, and n represents the number of data. When the skewness is a positive value, it indicates that the data distribution curve is skewed to the right compared to the curve of a normal distribution, that is, the tail part of the data distribution is on the right side of the mean; when the skewness is a negative value, it indicates that the data distribution curve is skewed to the left compared to the curve of a normal distribution, that is, the tail part of the data distribution is on the left side of the mean. When the skewness is equal to 0, it indicates that the distribution shape of the data is roughly the same as that of a normal distribution curve.

Coefficient of Variation: The coefficient of variation, also known as the “standard deviation rate,” is another statistical measure of the degree of variation of each observation in the data. The calculation formula is:

$$CV=S/\overline{X}$$

The coefficient of variation can eliminate the influence of different units and (or) average values on the comparison of the degree of variation of two or more data.

The calculation results of each index of the data are as follows.

| 变量名 | 样本量 | 最大值 | 最小值 | 平均值 | 标准差 | 中位数 | 方差 | 峰度 | 偏度 | 变异系数（CV） |
|---|---|---|---|---|---|---|---|---|---|---|
| ALTITUDE (1013) | 33498 | 3813 | 1 | 1306.621 | 566.155 | 1154 | 320531.205 | 7.695 | 2.306 | 0.43329687448010373 |
| Inertial Vertical Speed | 33498 | 2416 | 1 | 256.22 | 502.84 | 49 | 252848.49 | 4.614 | 2.352 | 1.962534431437682 |
| RADIO ALT | 33498 | 536 | 1 | 143.536 | 37.025 | 145 | 1370.868 | 38.144 | 2.633 | 0.257950753148906 |
| COMPUTED AIR SPD | 33498 | 945 | 1 | 344.969 | 127.402 | 291 | 16231.287 | 1.959 | 0.397 | 0.36931485264023534 |
| GROUNDSPEED | 33498 | 1436 | 1 | 784.279 | 186.375 | 796 | 34735.728 | 7.156 | -1.847 | 0.23763898977158884 |
| COG NORM ACCEL | 33498 | 86 | 1 | 12.943 | 7.446 | 11 | 55.442 | 20.072 | 3.196 | 0.5753046556734793 |
| PITCH ATT | 33498 | 305 | 1 | 215.168 | 74.686 | 243 | 5578.07 | 2.391 | -2.033 | 0.3471077720843552 |
| CAP CLM 1 POSN | 33498 | 149 | 1 | 4.597 | 9.405 | 3 | 88.456 | 72.661 | 7.709 | 2.0458704896342996 |
| ROLL ATT | 33498 | 415 | 1 | 23.814 | 46.742 | 12 | 2184.804 | 23.541 | 4.753 | 1.9628208775286815 |
| CAP WHL 1 POSN | 33498 | 330 | 1 | 68.78 | 49.527 | 62 | 2452.951 | -0.002 | 0.501 | 0.7200773267304699 |
| N1 SELTED-L | 33498 | 680 | 1 | 247.42 | 79.104 | 247 | 6257.48 | 6.41 | 0.897 | 0.3197169306585772 |
| N1 SELTED-R | 33498 | 678 | 1 | 248.728 | 79.024 | 246 | 6244.765 | 6.422 | 0.922 | 0.31771157895197577 |
| MAGNETIC HEADING | 33498 | 1569 | 1 | 714.165 | 327.852 | 660 | 107486.929 | -0.335 | -0.177 | 0.4590704140009221 |
| WIND DIR(ADIRU) | 33498 | 532 | 1 | 191.413 | 96.382 | 186 | 9289.43 | -0.339 | -0.15 | 0.5035282425816848 |
| WIND SPD(ADIRU) | 33498 | 711 | 1 | 265.764 | 136.008 | 254 | 18498.095 | 0.514 | 0.558 | 0.5117617965452251 |
| TRA-L | 33498 | 666 | 1 | 213.767 | 90.367 | 193 | 8166.24 | 6.592 | 2.077 | 0.422737446190112 |
| RUDD POSN | 33498 | 124 | 1 | 14.303 | 16.389 | 9 | 268.585 | 8.88 | 2.976 | 1.1457788849159336 |
| GLIDESLOPE DEV Dots (C) | 33498 | 968 | 1 | 332.481 | 227.092 | 361 | 51570.781 | -0.805 | 0.051 | 0.6830229169253369 |
| LOCALIZER DEV Dots (C) | 33498 | 1006 | 1 | 586.815 | 210.779 | 680 | 44427.705 | -0.339 | -0.887 | 0.3591909796581869 |
| PITCH ATT RATE | 33498 | 47 | 1 | 5.608 | 2.15 | 5 | 4.625 | 79.409 | 6.21 | 0.38345750618256813 |

From the table, it can be seen that among this batch of variables, by comparing the variance, kurtosis, skewness and coefficient of variation, it can be found that the coefficient of variation of Inertial Vertical Speed, CAP CLM 1 POSN, ROLL ATT, and RUDD POSN exceeds 1, indicating that these four variables have more outliers. Therefore, we have selected the box plots of these four variables as representatives, and the box plots of these four variables are shown below.

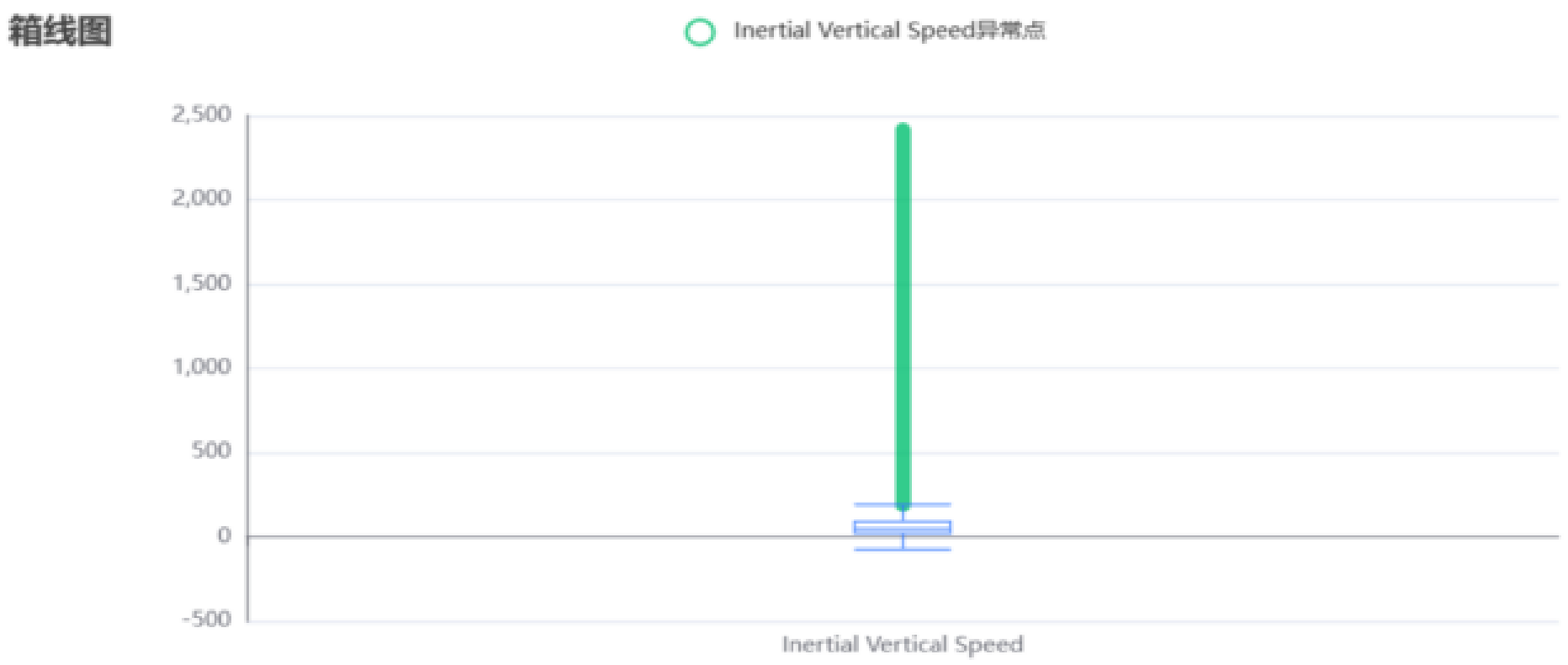

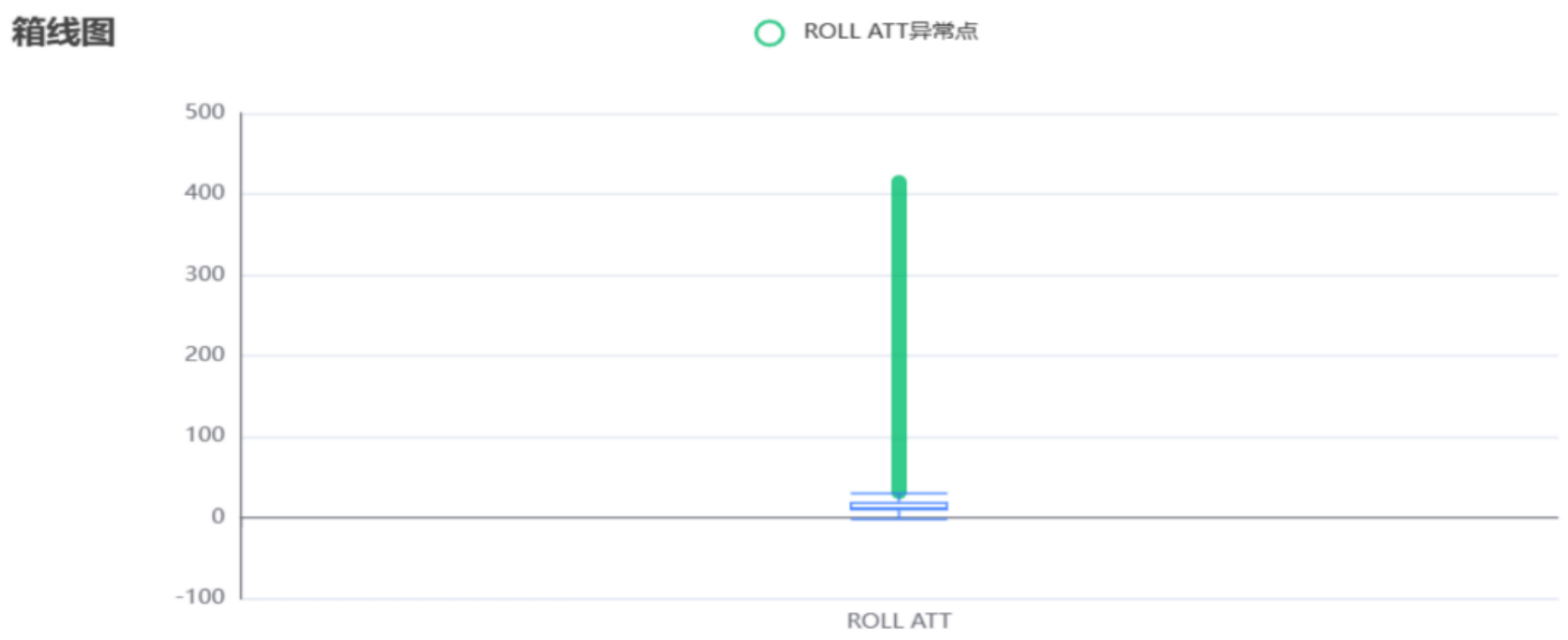

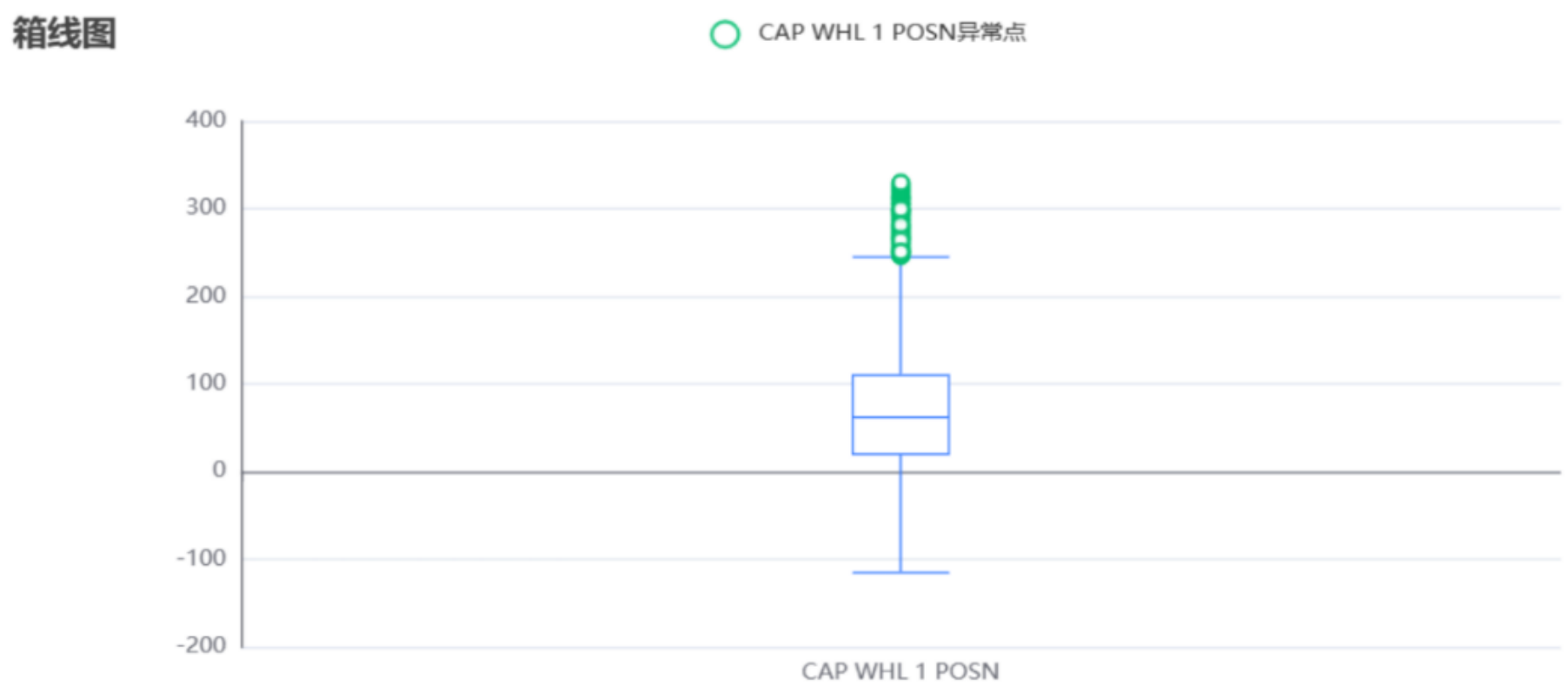

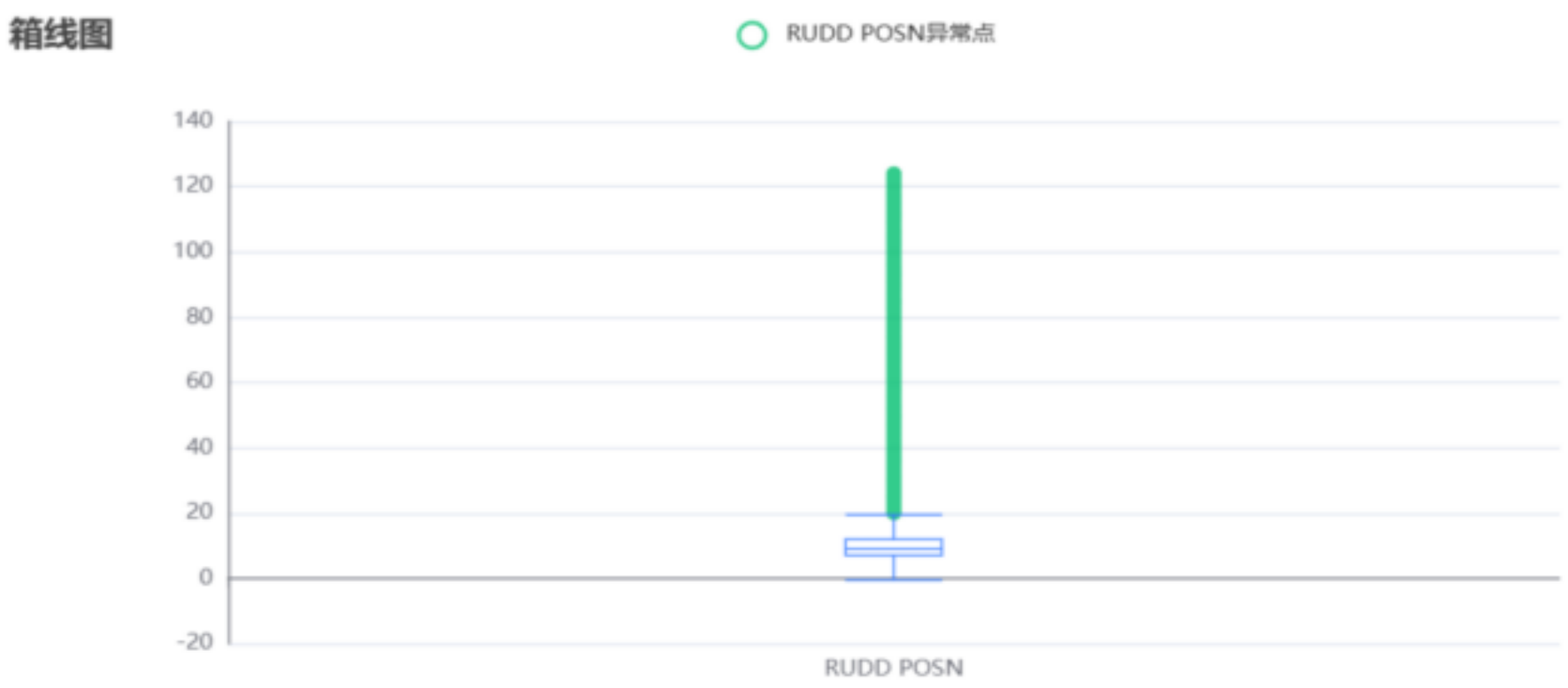

At the same time, as a comparison, we found that the coefficient of variation of GLIDESLOPE DEV Dots (C) is small, and the skewness and kurtosis are also small. The box plot of GLIDESLOPE

DEV Dots (C) is shown below.

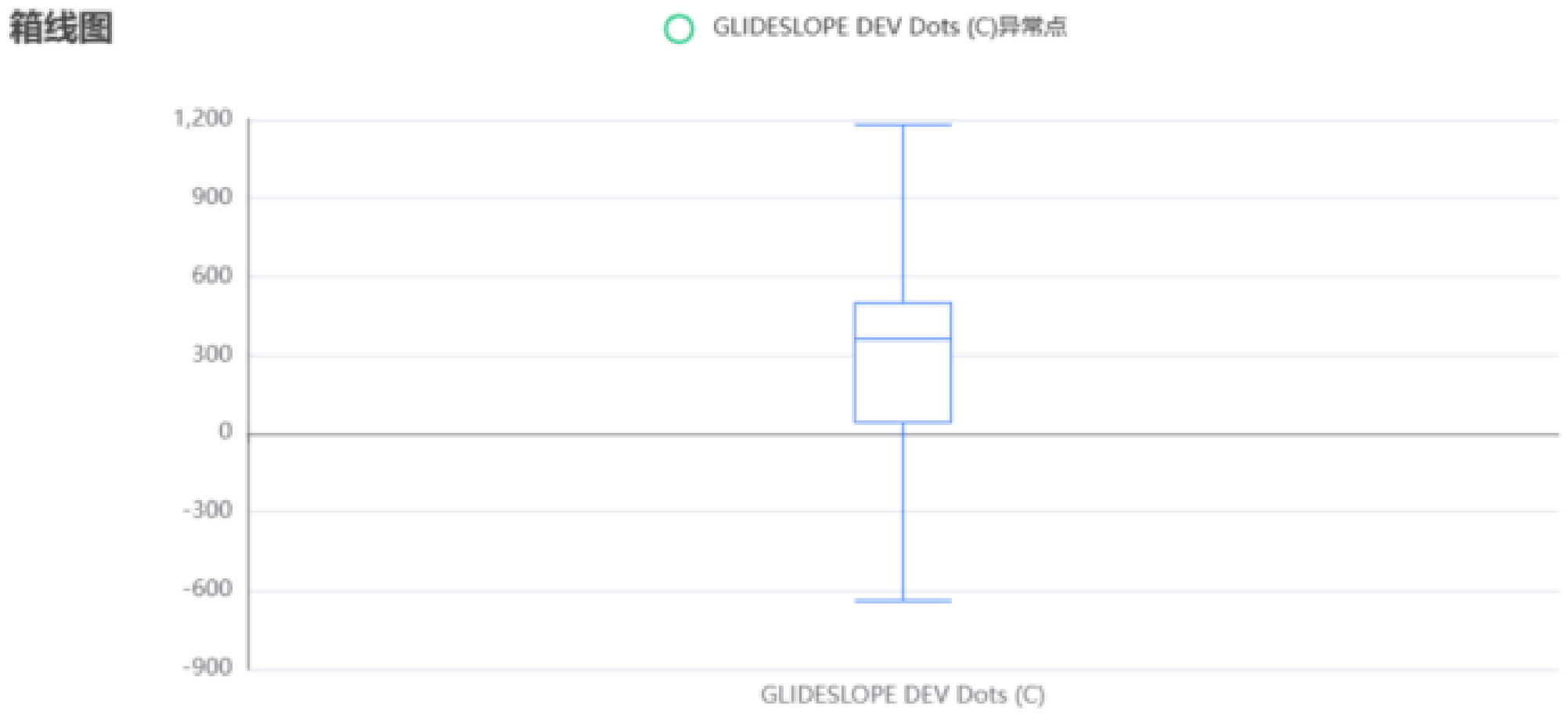

Therefore, we believe that the variance, kurtosis, skewness, and coefficient of variation can to some extent reflect whether there are outliers in the variables, and can be used to judge the reliability of the data. Through the analysis of each index of each variable above, we believe that the data reliability of Inertial Vertical Speed, CAP CLM 1 POSN, ROLL ATT, RUDD POSN, and TRA-L is insufficient and needs to be processed for outliers. The remaining data has a certain reliability, but in order to ensure the quality of the subsequent simulation, we have processed all the data for outliers.

## 3.4 Handling Outliers

### 3.4.1 Choosing a Handling Method

Question 1 clearly states that there are errors in the data. We first plot scatter plots of the unreliable data identified above to see their approximate trends. The plot is shown below.

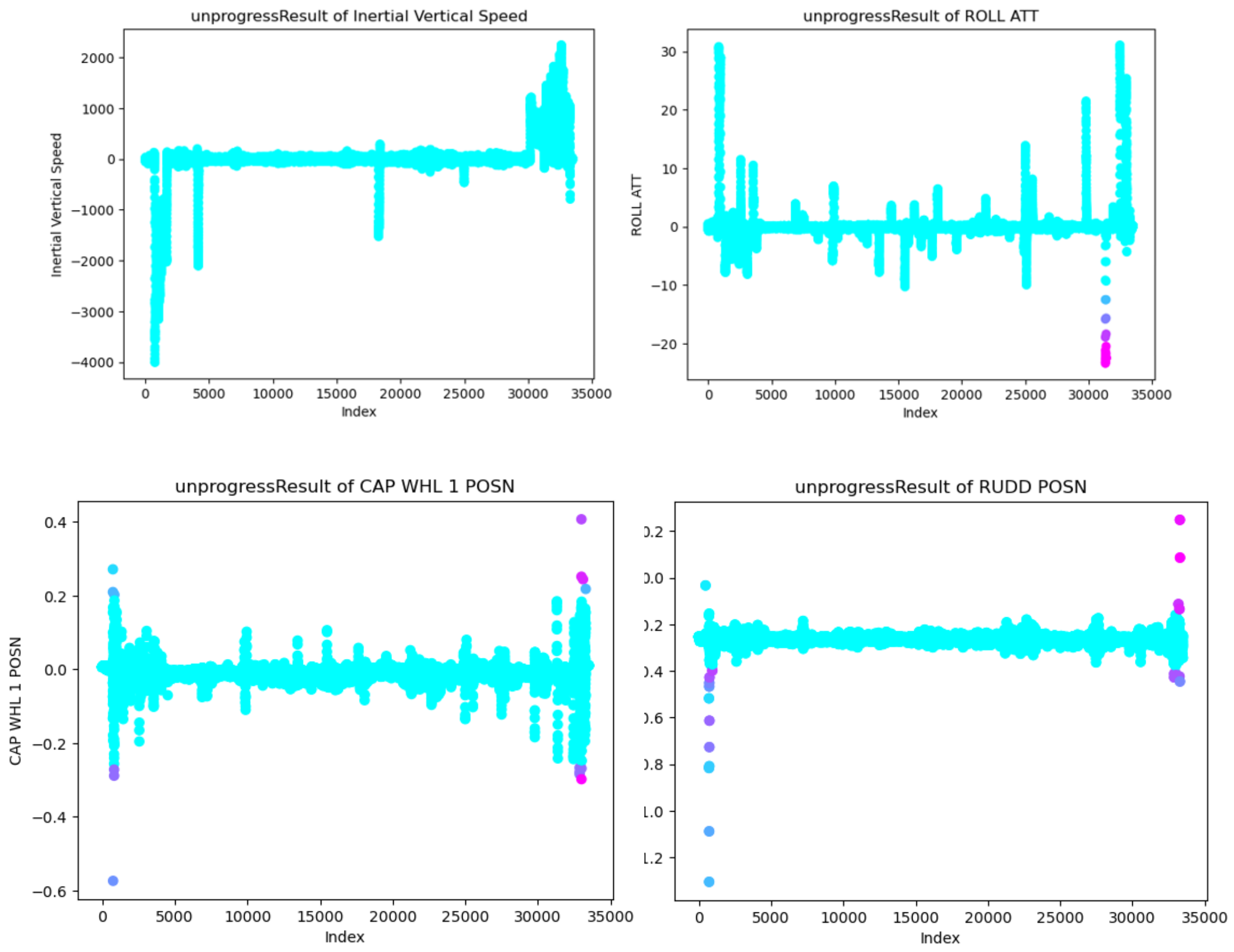

After plotting the scatter plot of the dataset, we found that the variables in the dataset generally show continuous changes or have a relatively stable trend. Therefore, we believe that outliers are points that are far from the trend of variable changes, that is, isolated points. Therefore, the point density at the location of the isolated point on the scatter plot will be low. In this question, since the shape and quantity of the scatter plot clustering are not fixed, we plan to use a density clustering algorithm to detect isolated points, that is, outliers.

Based on the density clustering DBSCAN (Density-Based Special Clustering of Applications with Noise), "density" can be understood as the tightness of sample points, and the measure of tightness requires evaluation using the radius and the minimum sample size. If the actual sample size exceeds the given minimum sample size threshold within the specified radius neighborhood, it is considered a high-density object. The DBSCAN density clustering algorithm can conveniently identify outliers in the sample set, so this algorithm is usually used to detect outliers.

In the DBSCAN clustering algorithm, points are divided into three categories.

Core point: points inside dense areas. If an object has more than Minpts number of points within its radius R, the object is a core point.

Border point: points on the edge of dense areas. If an object has a number of points less than MinPts within its radius R, but the object falls within the neighborhood of a core point, the object is a border point.

Noise point: points in sparse areas. If an object is neither a core point nor a border point, the object is a noise point.

Simply put, core points correspond to points inside dense areas, border points correspond to points on the edge of dense areas, and noise points correspond to points in sparse areas.

The algorithm is illustrated as follows.

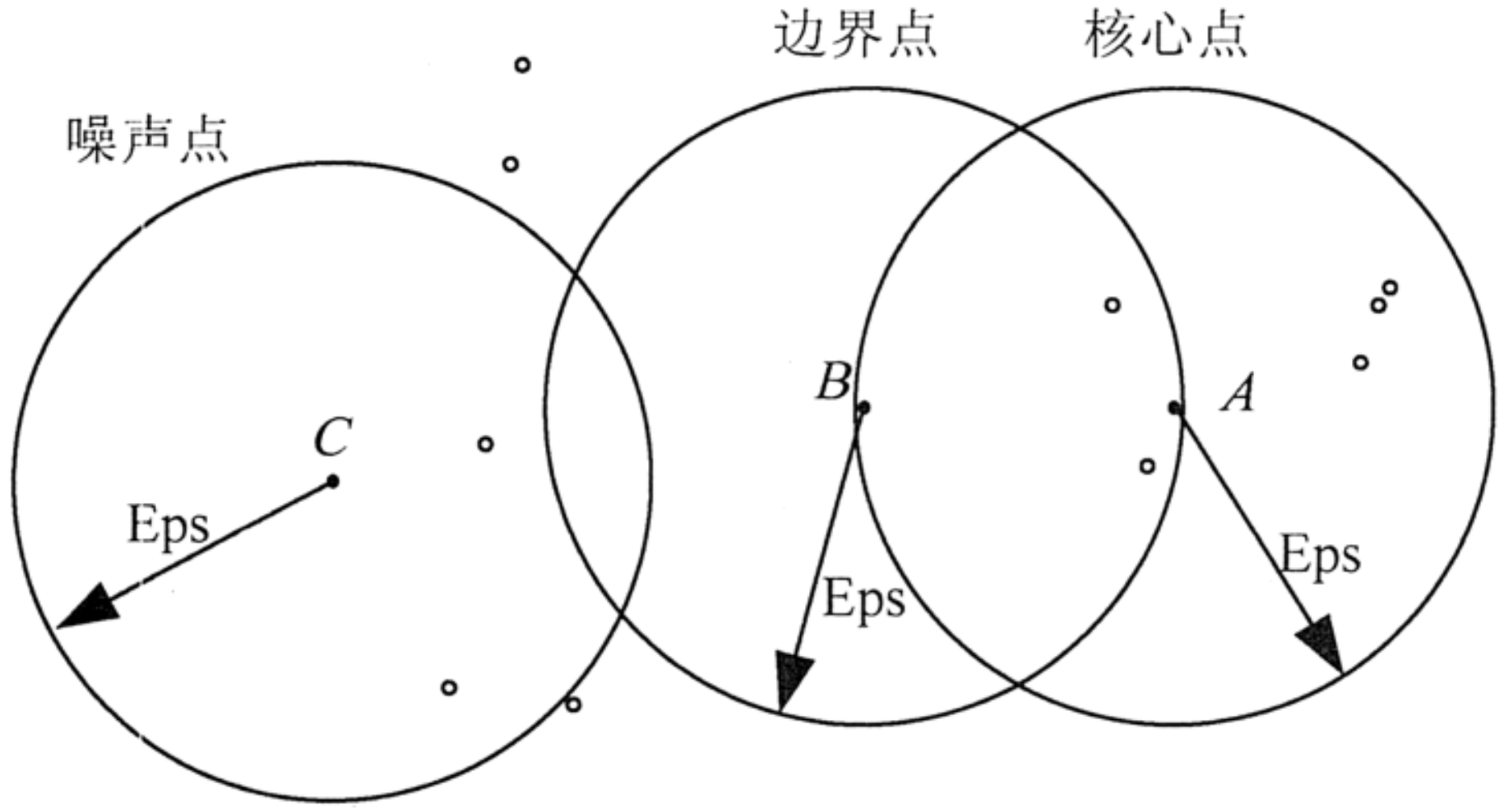

For this problem, we used the DBSCAN function in the sklearn.cluster module of python for clustering. The steps are as follows:

Input: set of samples $D = \{X_1, X_2, X_3 ... X_{N-1}, X_N\}$, neighborhood parameter $\{R, MinPts\}$

Initialize a set of core objects $\Omega$ and initialize the categories $k$.
Traverse the elements $D$. If it is a core object, add it to the set of core objects.
If all elements in the set of core objects have been visited, the algorithm ends. Otherwise, go to step 4.
In the set of core objects $D$, randomly select an unvisited core object $o$, mark it as visited, mark $o$ the category $k$, and then store $k$ the unvisited data in the $R$-neighborhood in the seed set.

If the seed set is empty, the current clustering cluster $C_K$ is generated, and $k = k + 1$,then jump to step 3. Otherwise, select a seed point from the seed set, mark it as visited and mark the category. Then, determine if it is a core object. If it is, add the unvisited seed point to the seed set and jump to step 5.

The output of the DBSCAN algorithm is the cluster label of each data point and the noise point label is -1. Therefore, we can obtain the location of the abnormal point.
If the density of the sample set is uneven and the distance between clusters is significantly different, the clustering quality is poor, and DBSCAN clustering is generally not suitable. In this problem, because there are many data and the dimensions of the data are significantly different, we need to first standardize the data. The steps are as follows:
Step 1: Read in the preprocessed dataset.
Step 2: Use scaler=StandardScaler() in python to standardize the data.
Step 3: Use the DBSCAN algorithm to identify abnormal values.
Step 4: Use the mean value to replace the abnormal values.
Step 5: Reverse standardize the data.
Finally, traverse all variables that need to process abnormal values, and export the results. The comparison before and after processing the abnormal values of the variables with low reliability, such as Inertial Vertical Speed, CAP CLM 1 POSN, ROLL ATT, and RUDD POSN, is shown below.

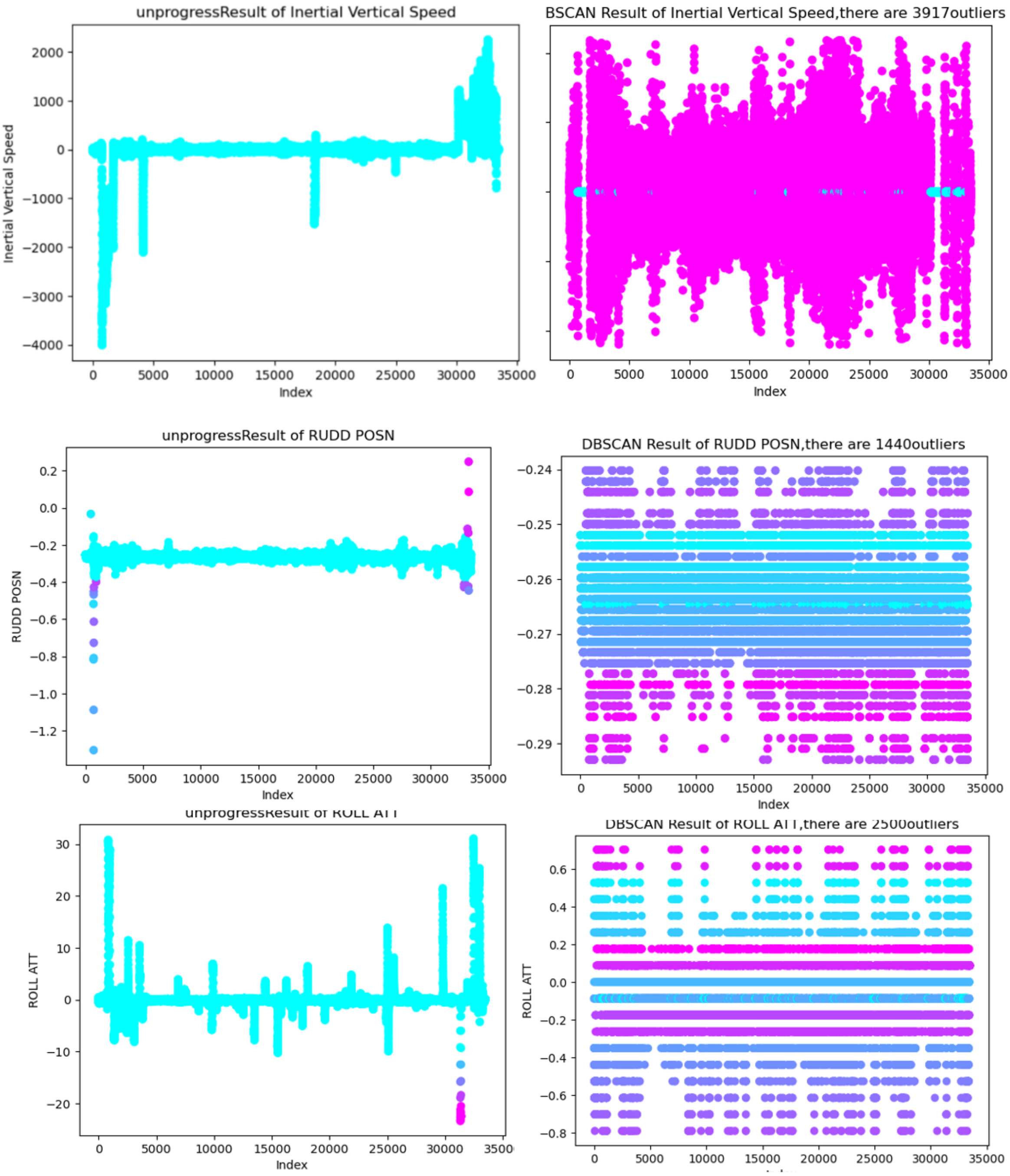
unprogressResult of Inertial Vertical Speed
Inertial Vertical Speed
Index
BSCAN Result of Inertial Vertical Speed,there are 3917outliers
Index
unprogressResult of RUDD POSN
RUDD POSN
Index
DBSCAN Result of RUDD POSN,there are 1440outliers
Index
ROLL ATT
Index
ROLL ATT

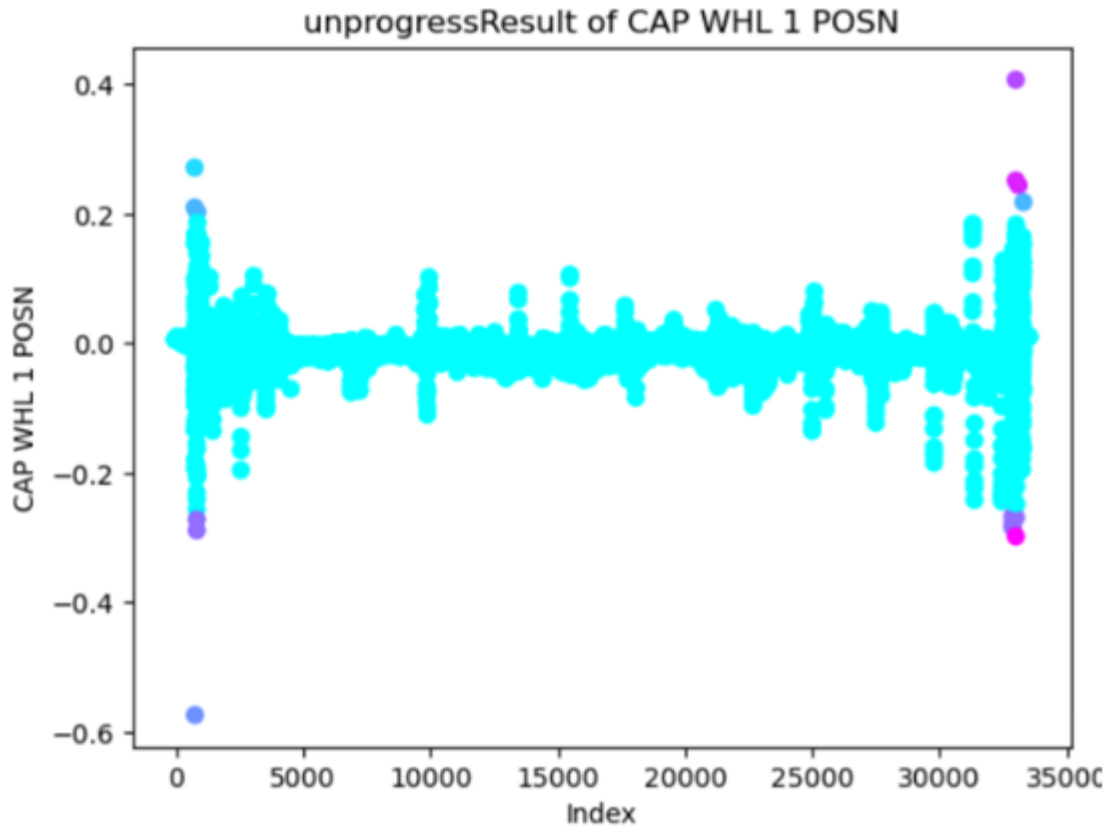

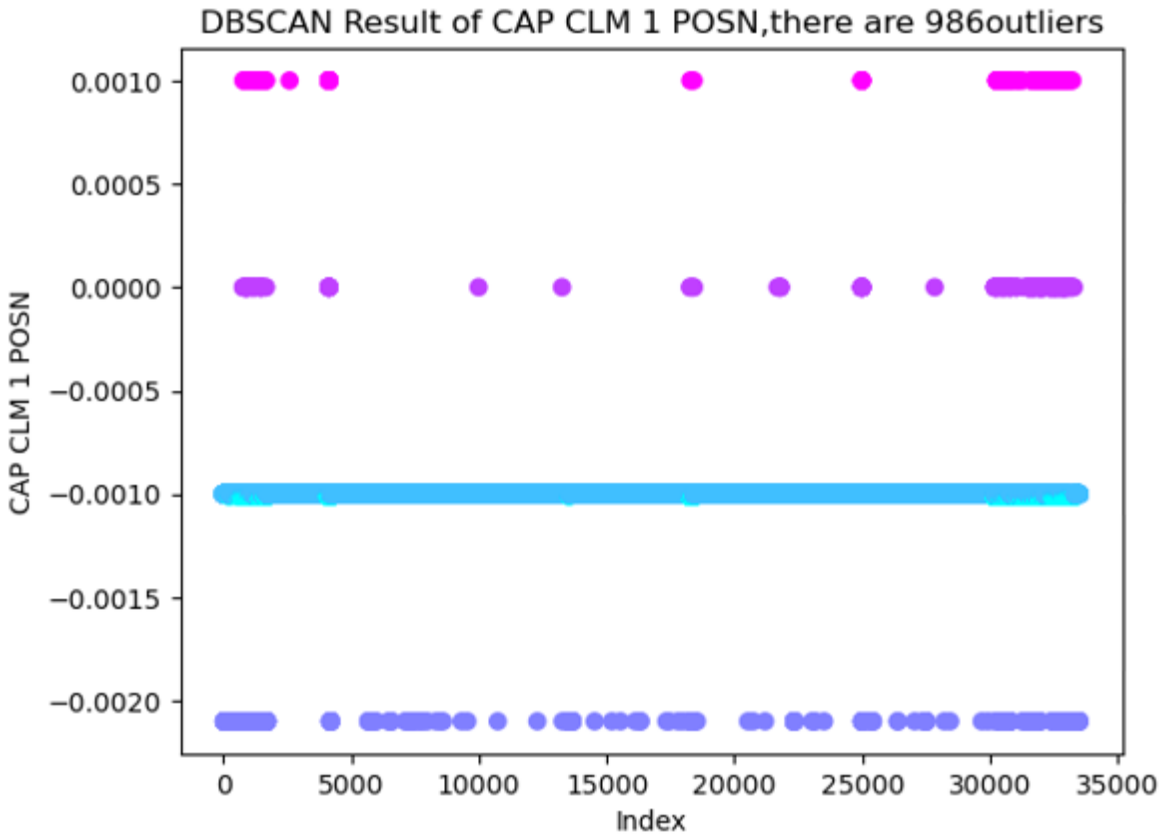

For variables with higher reliability, there is little change in their scatter plots after processing abnormal values. The following figure shows the comparison of the variable GROUNDSPEED before and after processing abnormal values.

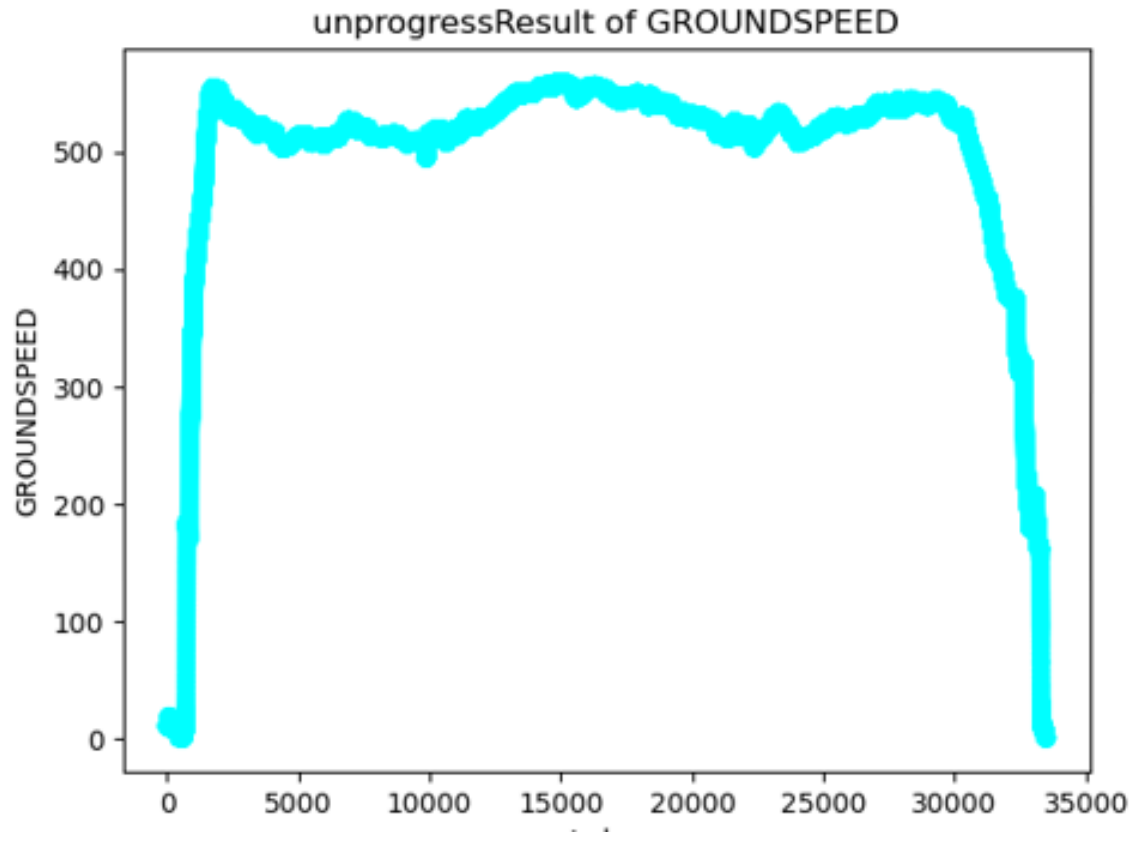

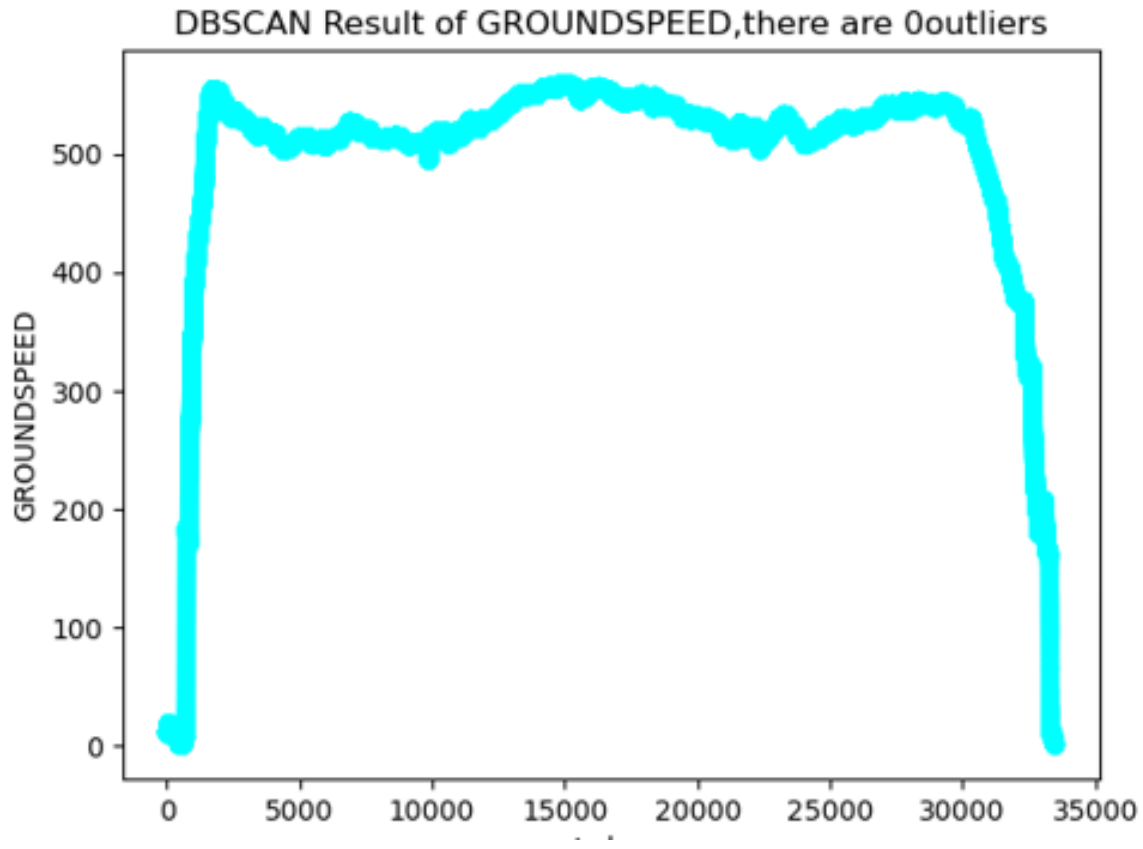

In summary, the DBSCAN algorithm processes abnormal values for variables with low reliability, but almost no processing for variables with high reliability.

## 3.5 Analysis of Key Data Items and Importance

Random Forest (RF) is a machine learning algorithm that uses multiple decision trees to train samples and integrate predictions. It is a classifier that uses bootstrap resampling techniques to randomly extract data from raw samples to construct multiple samples. Then, random splitting techniques are used to construct multiple decision trees for each resampled sample, and finally, multiple decision trees are combined and the final prediction result is obtained through voting.

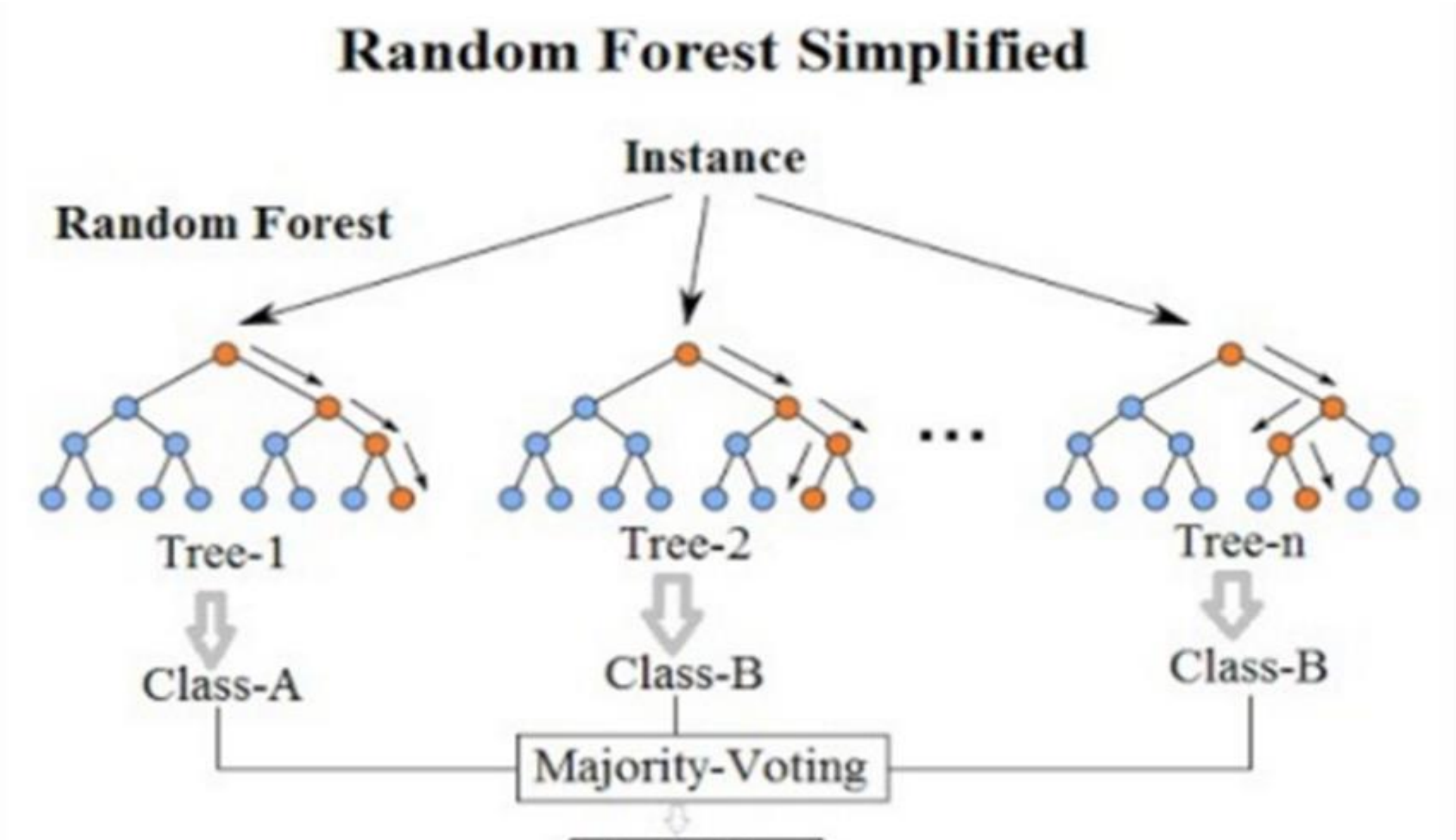

During the calculation process of a random forest, feature importance is determined by calculating the contribution of each feature to the model's predictive ability. This contribution is typically calculated based on the importance of the feature in each decision tree. Therefore, feature importance can be determined by calculating the number of nodes or the amount of impurity reduction that a feature is used for splitting in each decision tree. These values can be averaged across all decision trees to obtain the overall importance of the feature.

We used G-value as the target variable and other features as independent variables input into the random forest classifier to obtain key indicators and their weights. The feature importance is shown in the following figure:

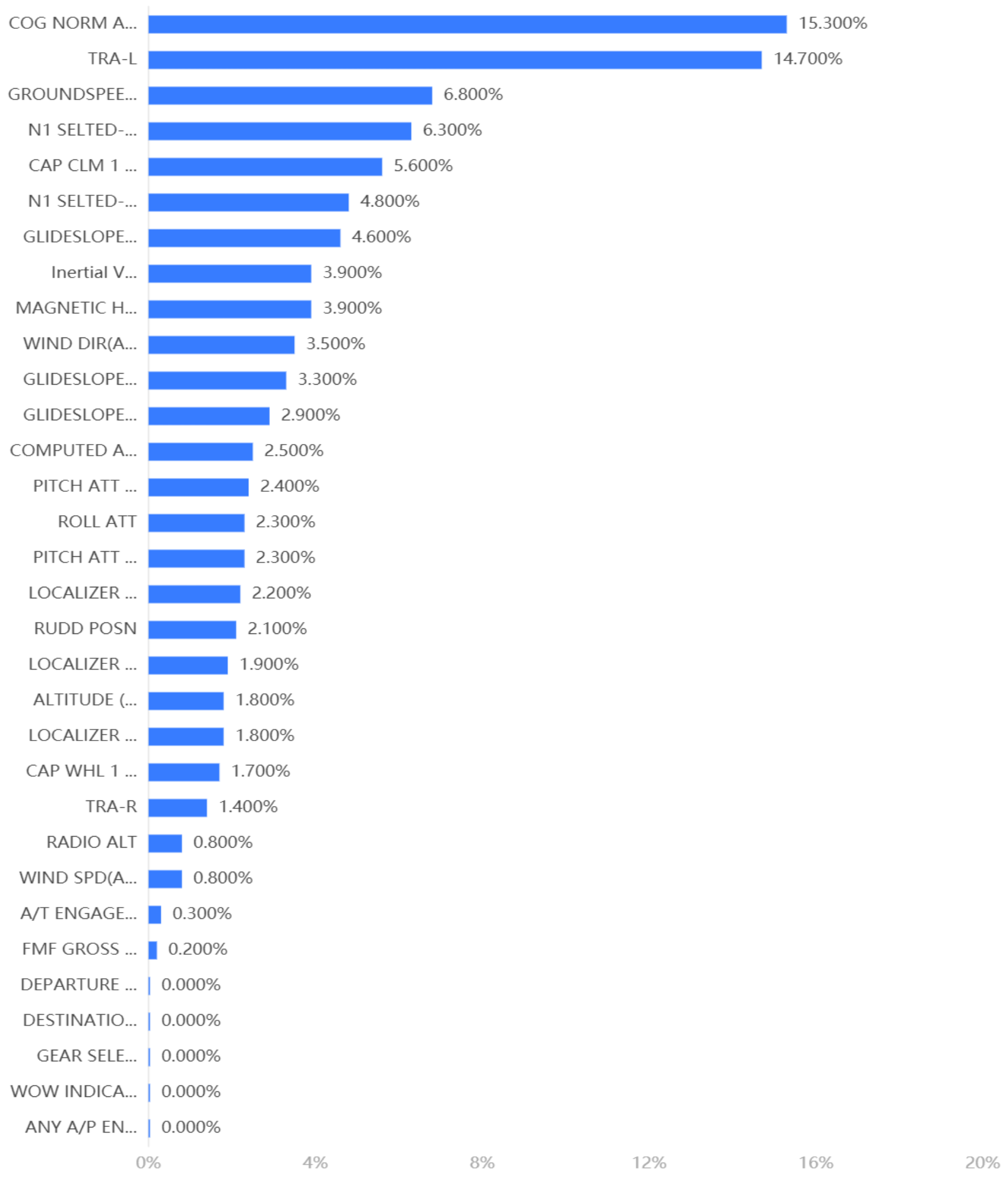

It can be seen that the most important indicator for flight safety model is COG NORM ACCEL, which is the landing G value, followed by TRA-R (right engine throttle lever position (angle)). Overall analysis shows that the stick quantity (CAP CLM 1 POSN) and disk quantity (CAP WHL 1 POSN) are also important influencing factors for flight safety.

At the same time, we also evaluated the random forest regression model and its inverse:

| | MSE | RMSE | MAE | MAPE |
|---|---|---|---|---|
| 训练集 | 26.826 | 5.179 | 4.253 | 35.882 |
| 测试集 | 114.702 | 10.71 | 8.154 | 46.355 |

The evaluation indicators include:
MSE (Mean Squared Error): The expected value of the square of the difference between the predicted value and the actual value. The smaller the value, the higher the accuracy of the model.
RMSE (Root Mean Squared Error): The square root of MSE. The smaller the value, the higher the accuracy of the model.
MAE (Mean Absolute Error): The average value of absolute error, which can reflect the actual situation of prediction error. The smaller the value, the higher the accuracy of the model.
MAPE (Mean Absolute Percentage Error): It is a percentage value of MAE. The smaller the value, the higher the accuracy of the model.

# 4. Model Building and Solving for Question 2

## 4.1 Problem Analysis

**Question 2** requires a reasonable quantitative description of flight control. Since the flight state is determined by the pilot's control, it is also possible to infer the flight control data from the flight state data. Therefore, we need to characterize and establish the mapping relationship between flight state data and control data. With rich data in Attachment 1, we chose the multi-input and multi-output BP neural network algorithm to characterize the mapping relationship more accurately to describe and predict state data using control data. First, it is necessary to process the data in Attachment 1 that has been cleared of outliers to obtain the mean and increase rate of relevant flight data. Then, the algorithm is used to train the data, and the flight control data that caused the flight state limit value at the limit time can be predicted. Finally, the model is reasonably tested by calculating the error to determine the rationality of the model.

## 4.2 Model Establishment

The aircraft control is mainly controlled by the central stick/wheel and throttle lever. The central stick/wheel controls attitude, which is located between the pilot's legs, and controls the aircraft's pitch, roll, and left/right rotation through hand operation. The side stick includes a spring and damping device and uses a fixed stick force curve. The pilot moves the handle forward and backward to achieve the aircraft's pitch and roll motion. At the same time, as the second sub-question of Question 1 shows, the throttle lever position is critical to flight safety.

### 4.2.1 Data Selection

As mentioned above, to obtain a reasonable quantitative description of aircraft control based on flight limit data and determine the cause of flight limit data based on aircraft control values, we need to choose the following data to characterize the relationship between flight state data and aircraft control data:
Flight state data: Landing G value (COG NORM ACCEL) Attitude (PITCH ATT)
Aircraft control data: Stick quantity (CAP CLM 1 POSN) Wheel quantity (CAP WHL 1 POSN) Left engine throttle lever position (TRA-L) Right engine throttle lever position (TRA-R)

## 4.2.2 BP Neural Network

To describe the relationship between the above data more objectively and reasonably, we introduce the BP neural network model using machine learning methods to explore and discover the inherent relationships between the data. The specific model theory is as follows [1][2]:
The BP neural network is a multi-layer feedforward neural network whose main feature is that the signal is forward propagated, and the error is backward propagated. Specifically, for the neural network model with only one hidden layer as follows:
The BP neural network process is mainly divided into two stages. The first stage is the forward propagation of the signal, which goes through the hidden layer and finally reaches the output layer. The second stage is the backward propagation of the error, from the output layer to the hidden layer, and finally to the input layer, adjusting the weights and biases from the hidden layer to the output layer and from the input layer to the hidden layer, respectively.

Example of a three-layer BP neural network

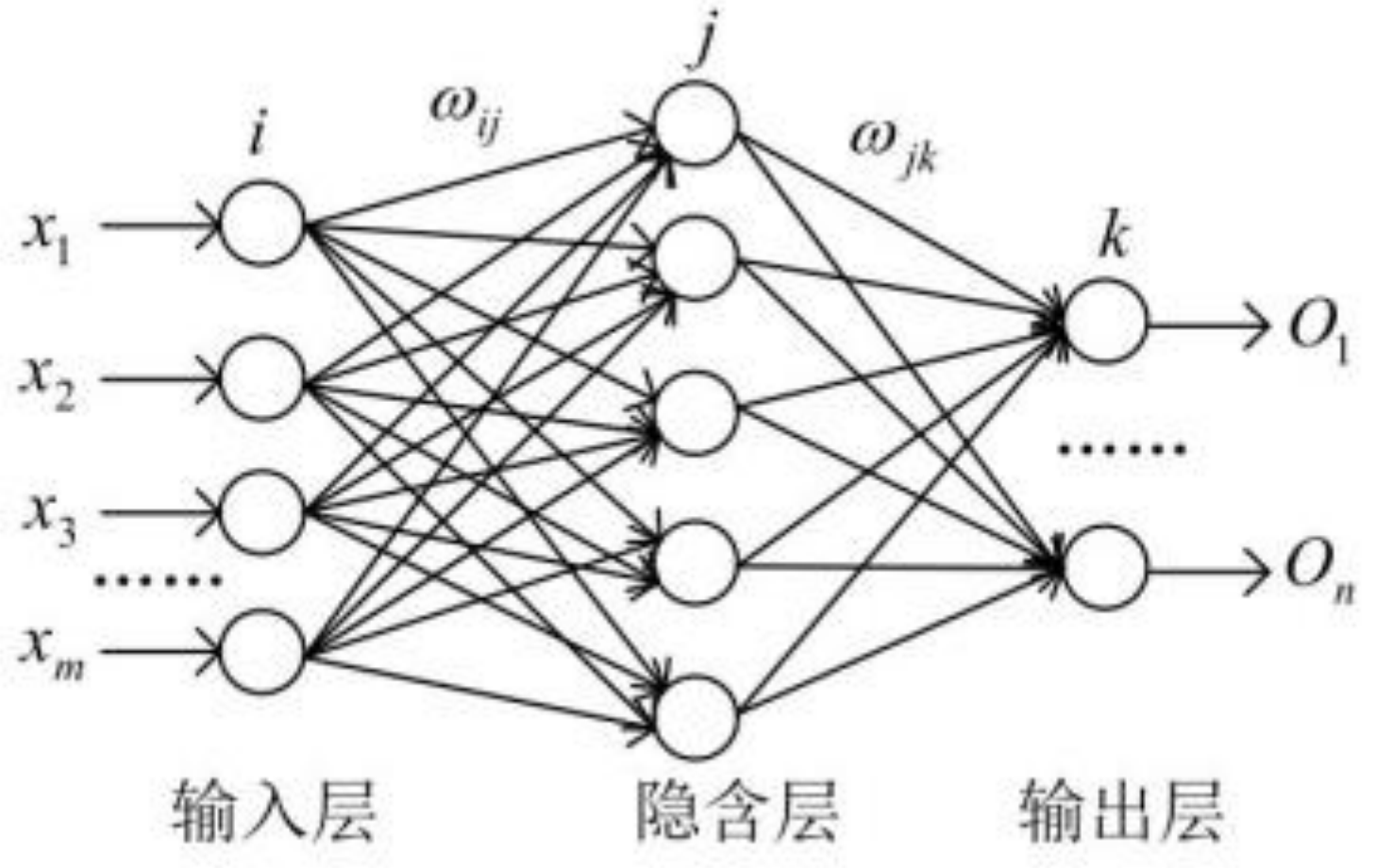

Let the output of the hidden layer be *Fj*, the output of the output layer be *Ok*, the activation function of the system be *G*, and the learning rate be *β*. Then there is the following mathematical relationship between the three layers:

$$F_j = G(\sum_{i=1}^{m} w_{ij}\, x_i + a_j)$$

$$O_k = \sum_{j=1}^{l} F_j\, w_{jk} + b_k$$

If the expected output of the system is *Tk*, then the system error *E* can be represented by the variance between the actual output value and the expected target value, as shown in the following expression:

$$E = \frac{1}{2}\sum_{k=1}^{n}(T_k - O_k)^2$$

And, using the principle of gradient descent, the update formulas for the system weights and

biases are as follows:

$$\begin{cases}\omega_{ij} = \omega_{ij} + \beta F_j(1-F_j)x_i \sum_{k=1}^{n} \omega_{jk} e_k \\ \omega_{jk} = \omega_{jk} + \beta F_j e_k \end{cases}$$

$$\begin{cases}a_j = a_j + \beta F_j(1-F_j)x_i \sum_{k=1}^{n} \omega_{jk} e_k \\ b_k = b_k + \beta e_k \end{cases}$$

## 4.3 Model Solution

### 4.3.1 Data preprocessing

This section takes attachment one table 201404070532 as an example.

Calculate the average value per second and growth rate of each selected data, and the results are shown in the attached table BP0532.

### 4.3.2 BP-Neural Network Training and Simulation

The training performance chart is as follows:

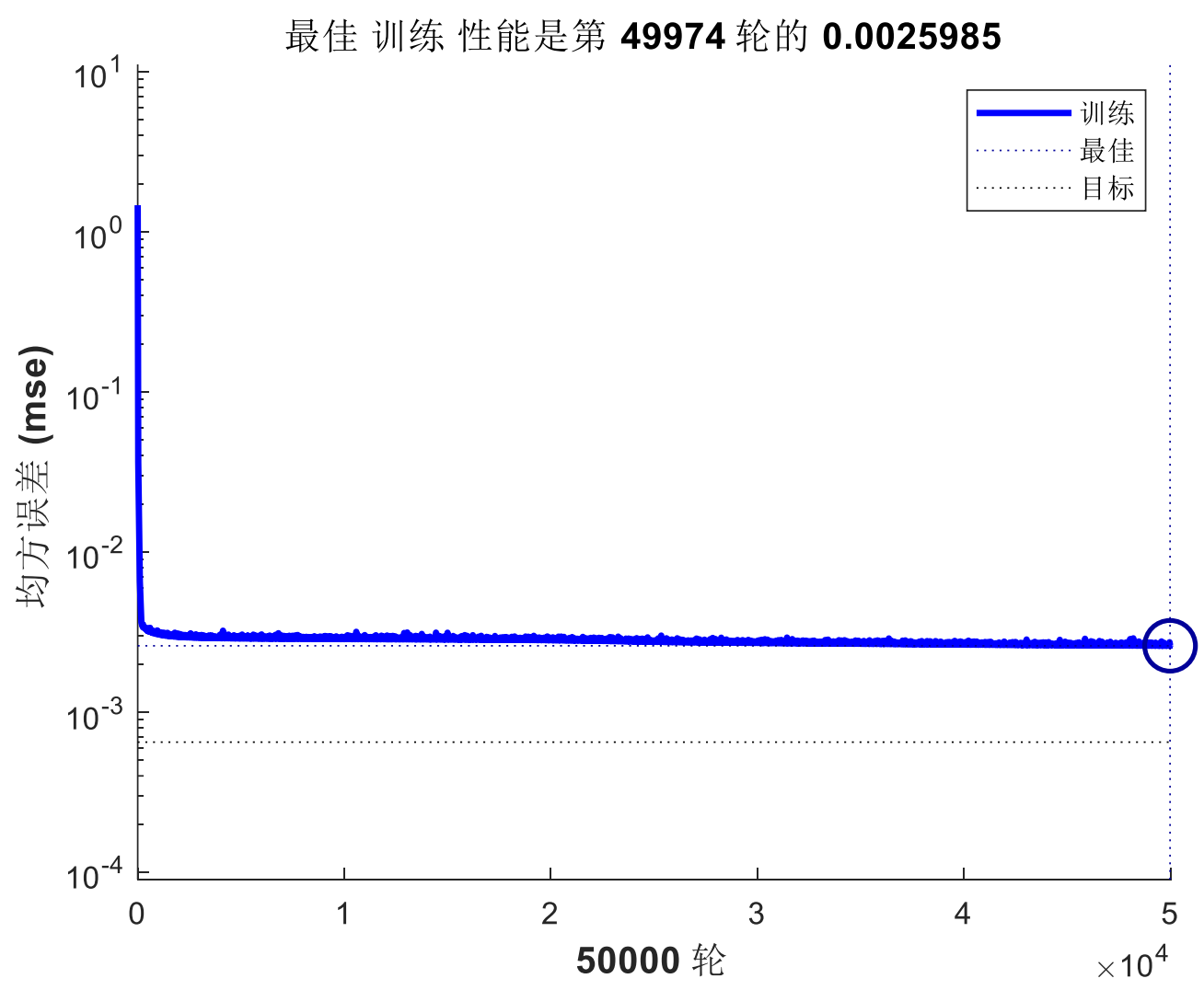

The training status is as follows:

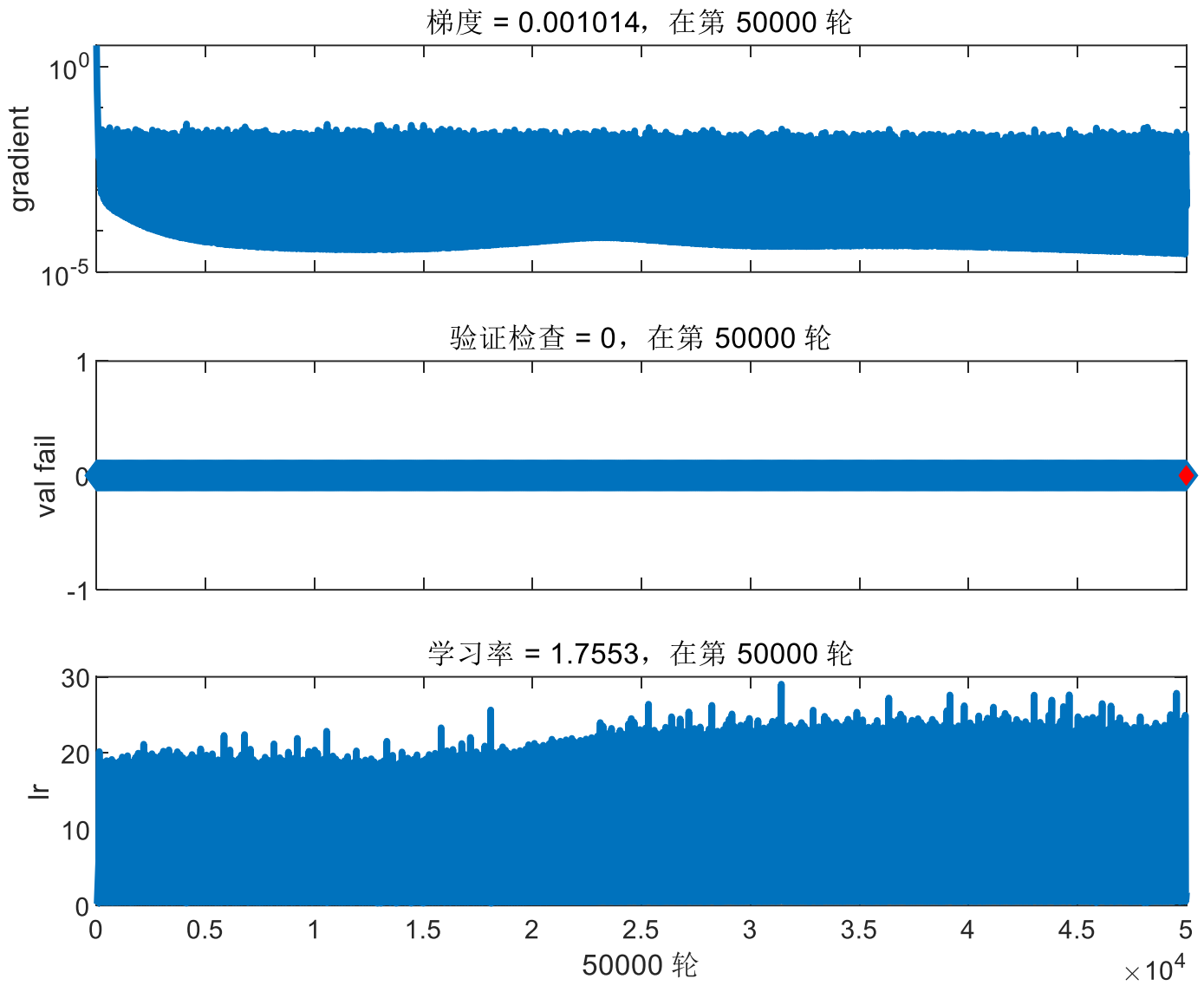

Training regression results are as follows:

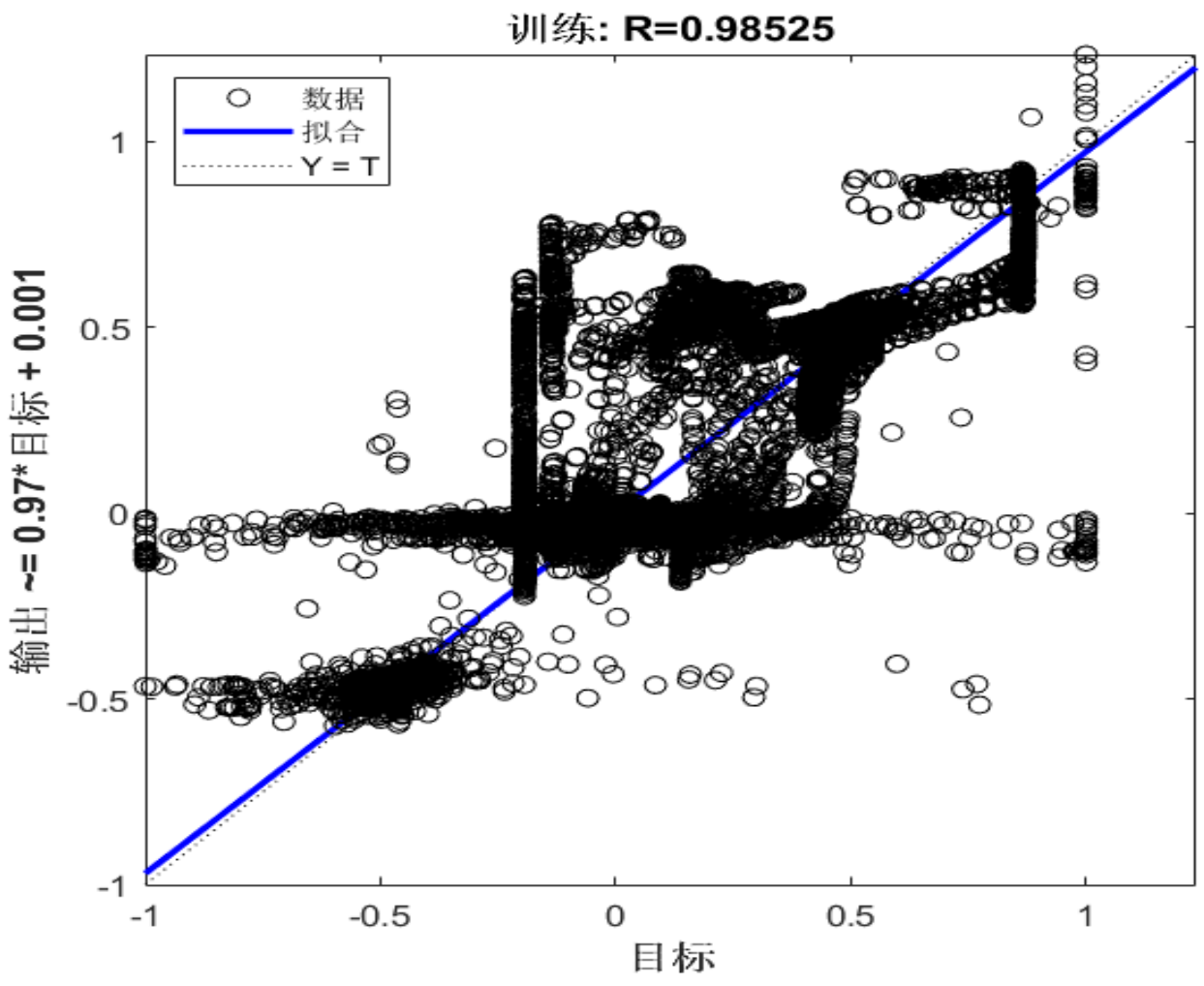

### 4.3.3 Rationality Test

The relative error of the predicted flight control data is shown in Table BP simulation result error. It can be seen that the relative error is small, indicating that the model is reasonable and can be used to characterize the relationship between the two.

# 5. Analysis of Problem Three's Excessive Data and Feature Exploration

## 5.1 Problem Analysis

According to the problem description, the causes of different excessive occurrences are different, and may be related to different airports, pilots, and routes. Annex 2 contains the occurrence of some excessive data from 2015 to 2017, including excessive names, takeoff airports, destination airports, and other data. Through the analysis of this table's excessive data, by studying the impact of different features on excessive data, using statistical methods to analyze the table's data through pivoting and cross-analysis, and thus studying the basic characteristics of different excessive.

In order to study the relationship between the occurrence of excessive events and factors, detailed investigation and analysis are needed.

After consulting relevant materials and papers, we have summarized four factors that may affect the occurrence of excessive events on airplanes, and the following are the impacts of these four factors:

(1) Human factors: excessive events may be related to the pilot's operation or decision-making, such as misjudgment, negligence or improper operation during takeoff or landing. In addition, factors such as pilot fatigue, psychological state, or health status may also affect their work performance.

(2) Technical factors: excessive events may be related to the design, maintenance, or failure of the aircraft, such as insufficient engine performance, on-board equipment failure or malfunction, etc. In addition, factors such as uneven aircraft load distribution and center of gravity deviation may also cause excessive events.

(3) Environmental factors: excessive events may be related to weather, airport conditions or air traffic control, such as strong winds, low visibility, snow and ice cover, etc., or environmental conditions such as insufficient runway length and high altitude.

(4) Management factors: excessive events may be related to the airline's operational management, training level, or regulations, such as insufficient pilot training, unreasonable flight plans or operating procedures, etc.

After analyzing the above factors and the importance of the data, we will conduct correlation analysis of excessive data from the following five aspects:

Explore the correlation between excessive types and warning levels

Explore the correlation between different aircraft numbers and excessive situations

Explore excessive situations on different routes

Explore the relationship between time and excessive events

Multi-dimensional factor analysis: explore the relationship between different aircraft numbers, routes, and excessive events

Analyze the distribution of routes and aircraft numbers under a certain excessive feature

## 5.2 Model Establishment and Data Analysis

### 5.2.1 Correlation between excessive types and warning levels

By extracting the EVENT_NAME (excessive name) from Annex 2, it can be found that there are 50 types of excessive names recorded in this table, which are:

```
array(['50英尺至接地距离远', 'GPWS警告(sink rate)', '下降率大400-50ft', '接地速度小',
       '下降率大2000-1000（含）ft', '进近坡度大50ft以下', '爬升速度大35-1000ft', '着陆俯仰角小',
       '转弯滑行速度大', '空中过载', '着陆垂直载荷大', '进近速度大50 ft以下', '低于下滑道', '高于下滑道',
       '进近速度大500-50（含）ft', '下降率大1000-400(含)ft', '直线滑行速度大',
       '坡度大400/1500ft以上', '离地速度大', '爬升速度小35-1000ft', '起飞坡度大0-35（含）ft',
       '进近坡度大200-50（含）ft', '未知', '抬前轮速度小', '收反推晚', '放起落架晚', '抬前轮速度大',
       '抬头速率小', '低空大速度2500ft以下', '下滑道警告', '进近坡度大1500-500（含）ft',
       '进近坡度大500-200（含）ft', '着陆襟翼到位晚', '抬头速率大', 'GPWS警告(windshear)',
       'High normal accel with flap (in flight)', '着陆重、跳着陆', '50英尺至接地距离近',
       '着陆俯仰角大', '起飞收起落架晚', '起飞EGT超限', '超襟翼限制速度Vfe', '复飞', '离地仰角大',
       'Left of centreline below 1000ft', 'TCAS RA 警告', '离地速度小',
       '爬升坡度大35-400(含)ft', '进近速度小500 ft以下', '起飞滑跑方向不稳定'], dtype=object)
```

Count the frequency of various types of exceeded limitations and draw a distribution chart as follows:

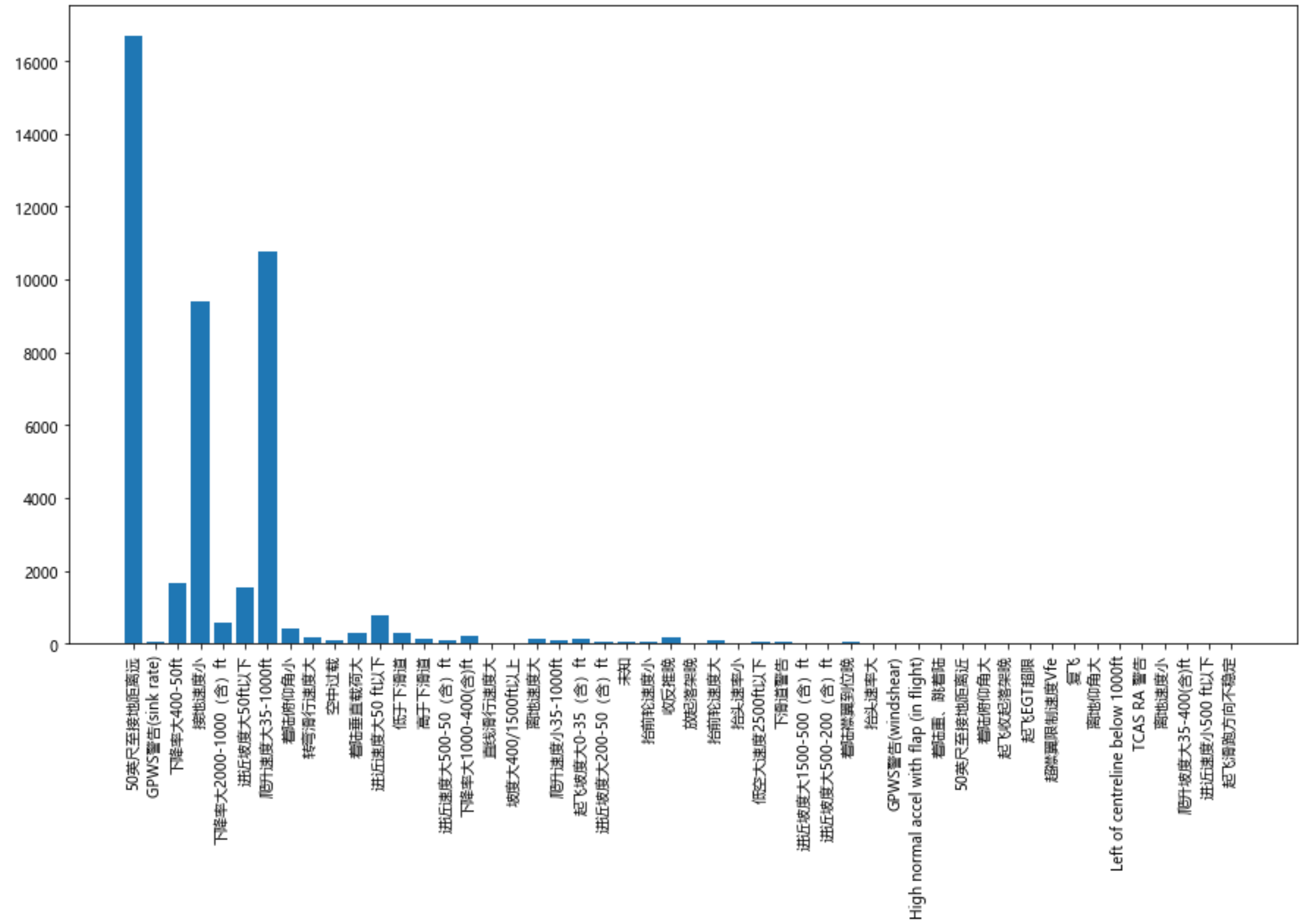

It can be seen that the occurrence rate of three types of over-limit events is much higher than that of other over-limit events:

| | 超限类别 | 超限次数 |
|---|---|---|
| 0 | 50英尺至接地距离远 | 16692 |
| 1 | 爬升速度大35-1000ft | 10748 |
| 2 | 接地速度小 | 9399 |

Among them, the occurrence rate of "50 feet to ground distance far" event accounts for 38% of the total over-limit events, and it occurs extremely frequently.

By studying the "ALERT (warning level)" column in the data, it can be found that there are two types of over-limit event classifications given in the data, level 2 and level 3. Through comparative research, it can be found that the same over-limit name may correspond to different over-limit levels, which represents the severity of a certain situation.

Therefore, it is difficult to conduct correlation analysis between these two, so we explored the correlation between warning level and overall over-limit situation.

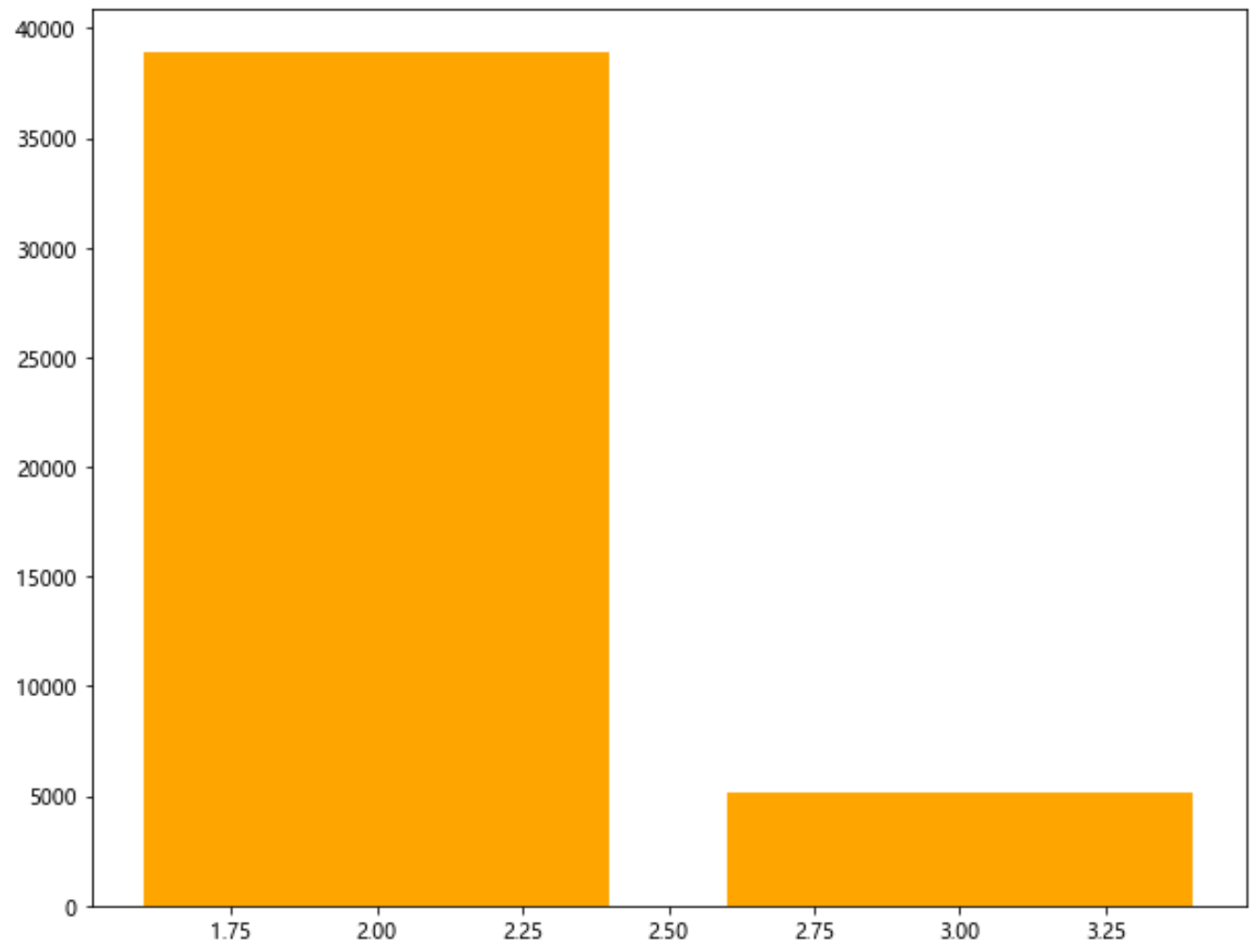

By drawing the distribution of second-level and third-level exceedances, it can be concluded that the number of second-level exceedances is almost eight times that of third-level exceedances, and a large part of second-level exceedances occur when the distance is "50 feet to ground distance far". This exceedance occurs during the landing phase of aircraft flight, so it can be judged that: **the landing phase is a period with a high incidence of aircraft exceedance events, and data from the landing phase should be measured and prevented with emphasis**.

### 5.2.2 The correlation between different aircraft numbers and exceedance events

The distribution of the number of exceedances for different aircraft numbers was plotted by taking the ARN (aircraft number) column from the data, as shown below:
The aircraft numbers and their corresponding number of exceedances for the top ten aircraft with exceedance events are as follows:

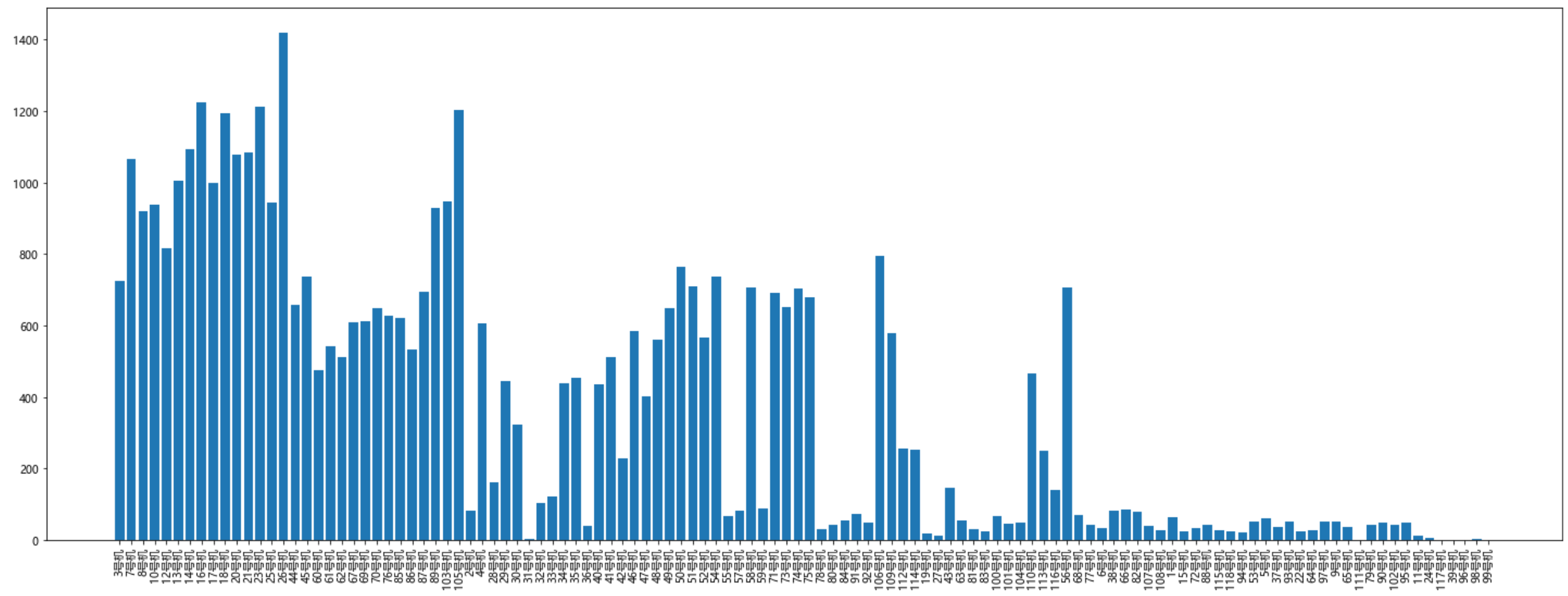

The aircraft registration numbers and their corresponding number of exceeded limits for the top ten aircraft with exceeded limits events are as follows:

| | ARN | 超限次数 |
|---|---|---|
| 0 | 26号机 | 1419 |
| 1 | 16号机 | 1225 |
| 2 | 23号机 | 1211 |
| 3 | 105号机 | 1202 |
| 4 | 18号机 | 1194 |
| 5 | 14号机 | 1092 |
| 6 | 21号机 | 1083 |
| 7 | 20号机 | 1078 |
| 8 | 7号机 | 1067 |
| 9 | 13号机 | 1004 |

Therefore, airlines can inspect and maintain the aircraft that have had frequent occurrences of exceeding limits, and determine whether it is a hardware issue with the aircraft.

### 5.2.3 Investigating the Occurrence of Overlimit Situations on Different Routes

Table 2 in the attachment provides information on both ARR (destination airport) and DEP (departure airport), which allows us to determine the relationship between route data and overlimit situations.
It is common knowledge that ARR-DEP and DEP-ARR are the same route. Therefore, by processing and deduplicating the data for destination and departure airports, we can obtain route data. The distribution of overlimit incidents on different routes can be plotted as follows:
Eg: Use 3-68 to represent the route between Airport 3 and Airport 68.

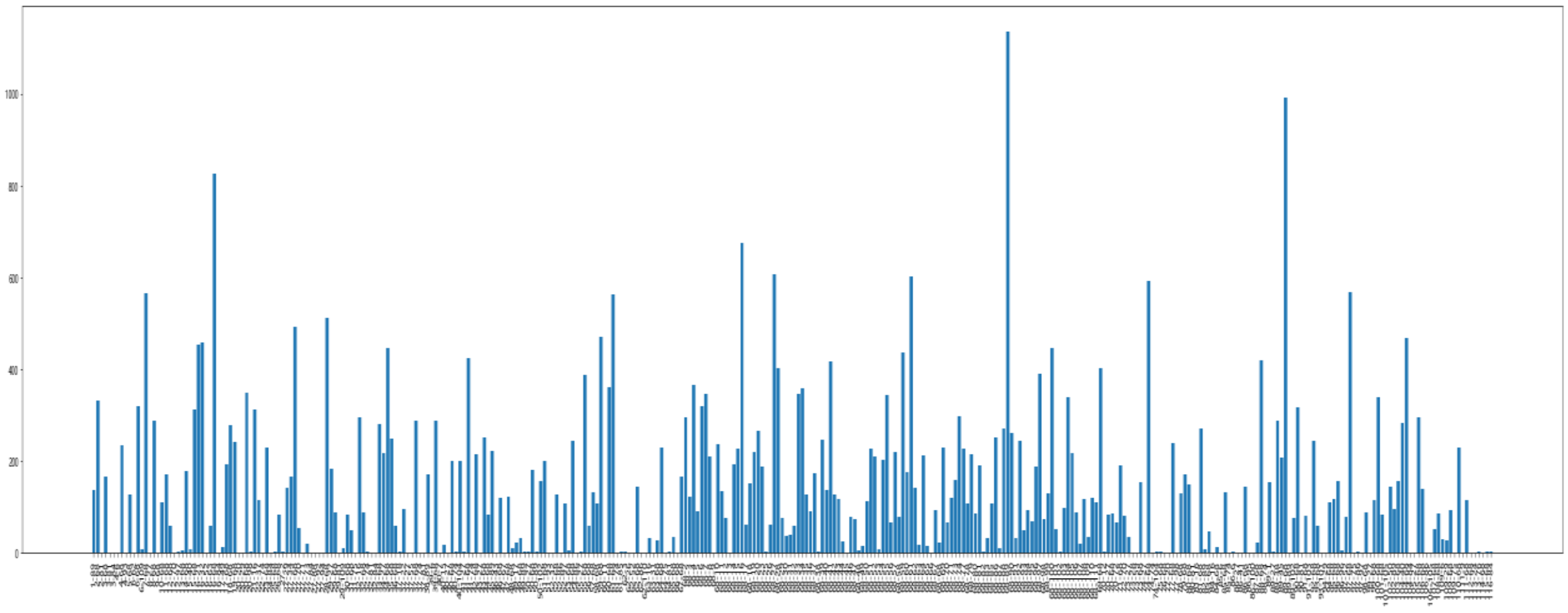

The aircraft registration numbers and their corresponding number of exceeded limits for the top ten aircraft with exceeded limits events are as follows:

| | ARN | 超限次数 |
|---|---|---|
| 0 | 26号机 | 1419 |
| 1 | 16号机 | 1225 |
| 2 | 23号机 | 1211 |
| 3 | 105号机 | 1202 |
| 4 | 18号机 | 1194 |
| 5 | 14号机 | 1092 |
| 6 | 21号机 | 1083 |
| 7 | 20号机 | 1078 |
| 8 | 7号机 | 1067 |
| 9 | 13号机 | 1004 |

For these ten routes, an analysis can be conducted on factors such as airport runways, signal towers, and command personnel to identify the reasons for the excessive number of limit violations.

### 5.2.4 Investigating the Relationship between Time and Limit Violations

5.2.4.1 Investigating the Impact of Month on Limit Violations

Extracting the date data from the table and taking out the month data, it can be found that the distribution of limit violations is very uniform and has no obvious correlation with the month.

5.2.4.2 Investigating the Impact of Working Days on Limit Violations

Extracting the date data from the table, using the datetime.strptime() function to convert the string data to datetime data, and then using the weekday() function to determine whether it is a working day or a weekend.

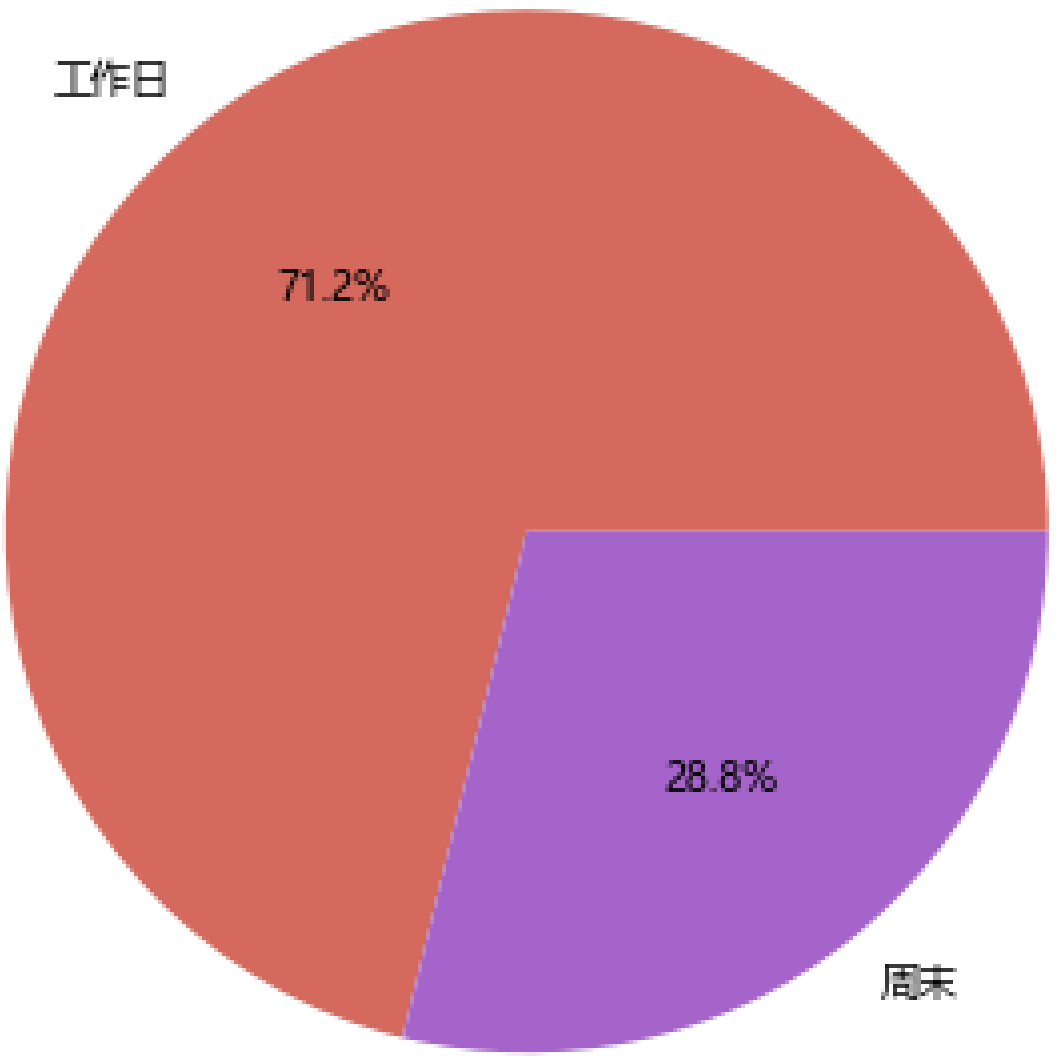

Based on the analysis, it can be concluded that there is no significant correlation between the occurrence of overlimit incidents and whether it is a workday or not, with a ratio of approximately 71.2%:28.8% for workdays and weekends, which is consistent with the division of five working days and two rest days.

### 5.2.5 Analysis of the Relationship between Different Aircraft Numbers, Routes, and Overlimit Incidents

We attempted to conduct multidimensional analysis by simultaneously analyzing multiple factors to explore the correlation between overlimit incidents and aircraft numbers and routes.
Firstly, we encoded the overlimit incidents and routes, and then drew a 3D scatter plot of aircraft numbers, routes, and overlimit incidents:

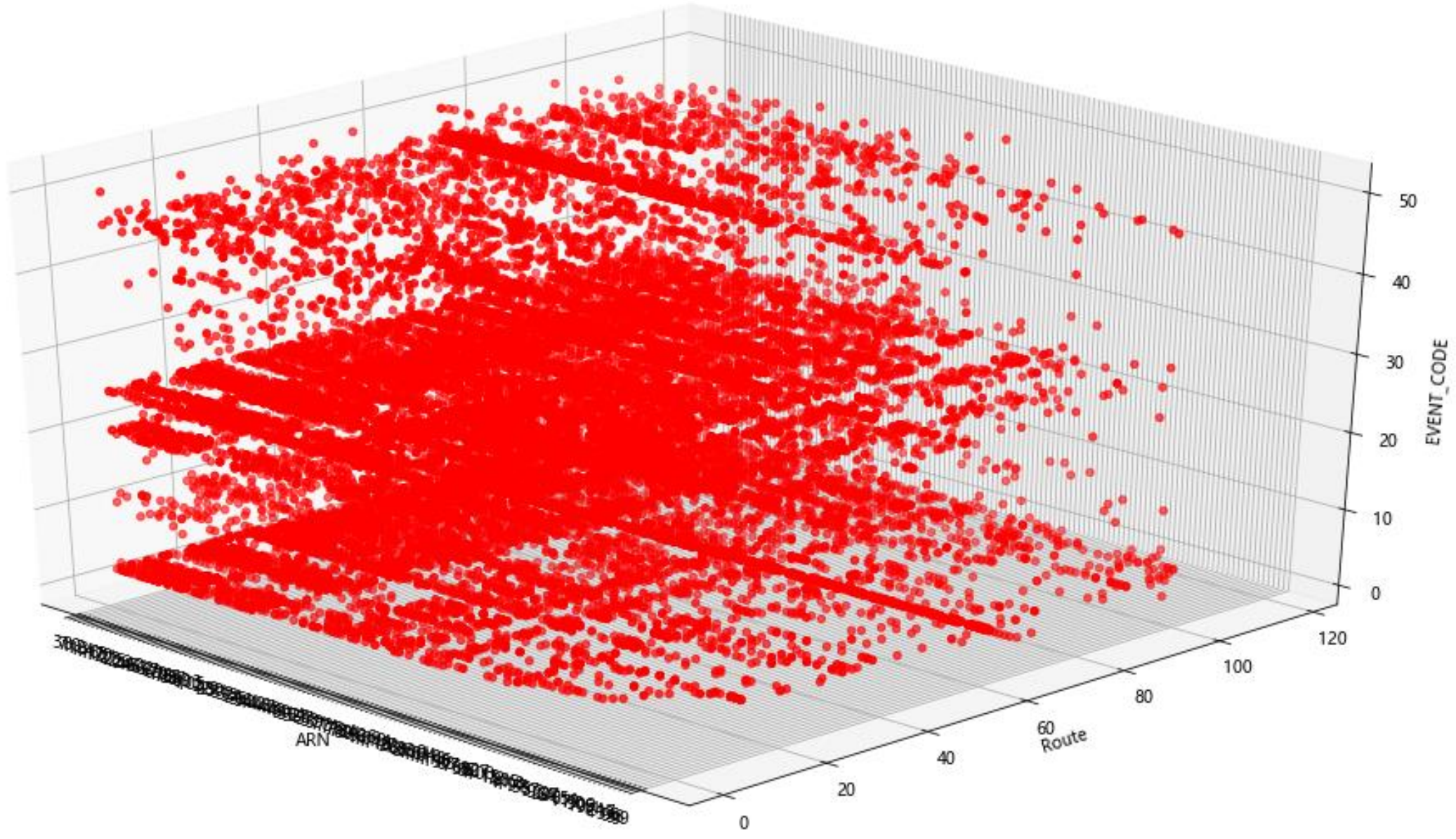

It can be observed that the 3D image shows a layered phenomenon, indicating that certain exceedance events have a higher probability of occurrence, while the distribution on the ARR axis is relatively uniform.

### 5.2.6 Analyzing the distribution of flight routes and aircraft numbers under a certain exceedance feature

In the above analysis, we explored the factors that cause exceedance features based on the features themselves. In this analysis, we focus on a specific exceedance situation to analyze the distribution of flight routes and aircraft numbers.
As a large portion of the exceedance features involve "50 feet to touchdown distance far," we choose this exceedance situation to analyze the distribution of aircraft numbers and flight routes.
5.2.6.1 Distribution of Aircraft Numbers

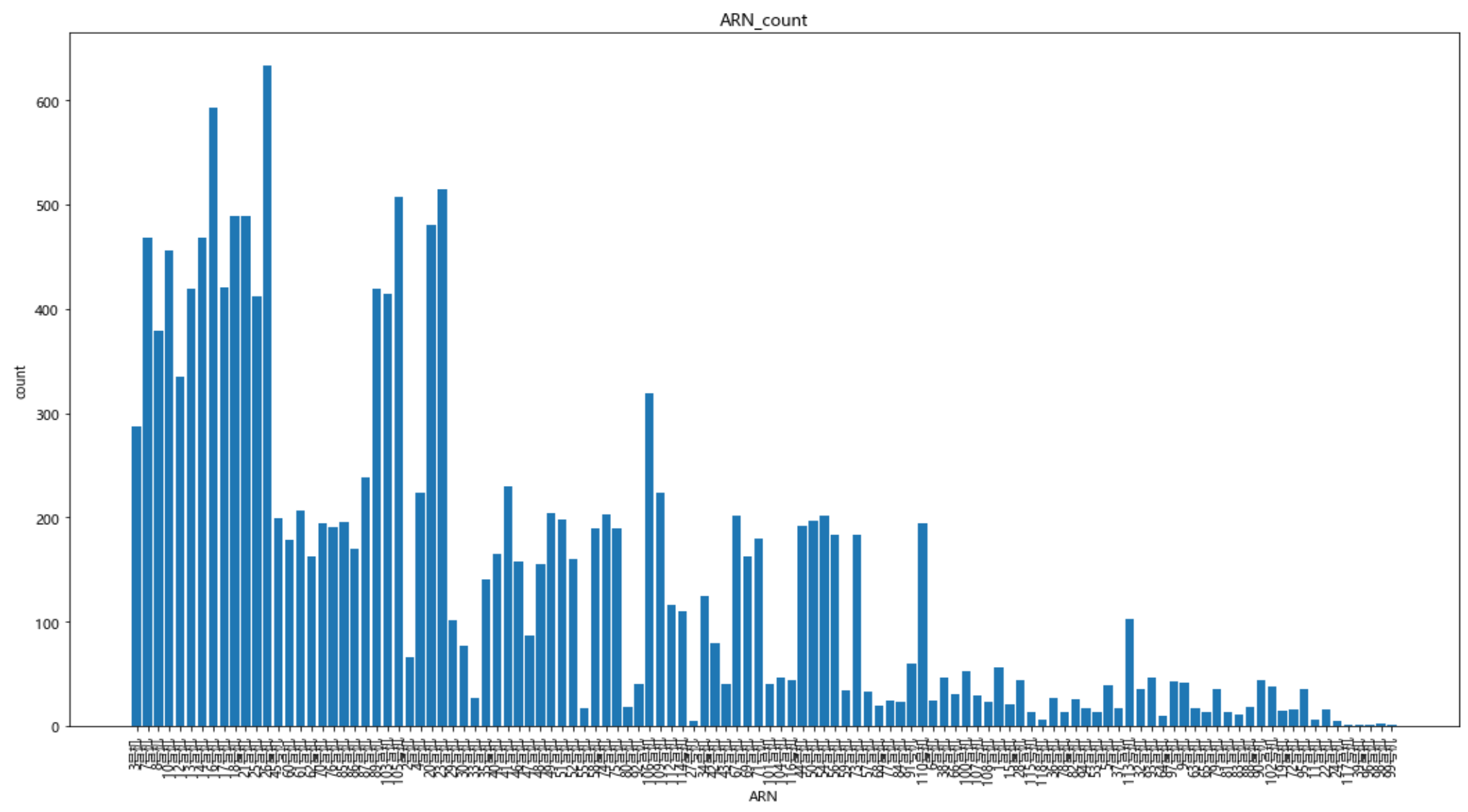

It can be seen that after extracting the over-limit situation of "50 feet to the ground distance", the distribution of ARN is basically consistent with using all over-limit situations.
The top ten aircraft with the most over-limit situations are:

```
['26号机',
 '16号机',
 '23号机',
 '105号机',
 '18号机',
 '21号机',
 '20号机',
 '7号机',
 '14号机',
 '10号机']
```

It is also consistent with the use of the overall over-limit situation analysis.
5.2.6.2 Distribution of routes
When using the overall over-limit situation analysis, we use route data, which integrates the destination airport and the departure airport. In this analysis, we can separate DEP and ARR for analysis.

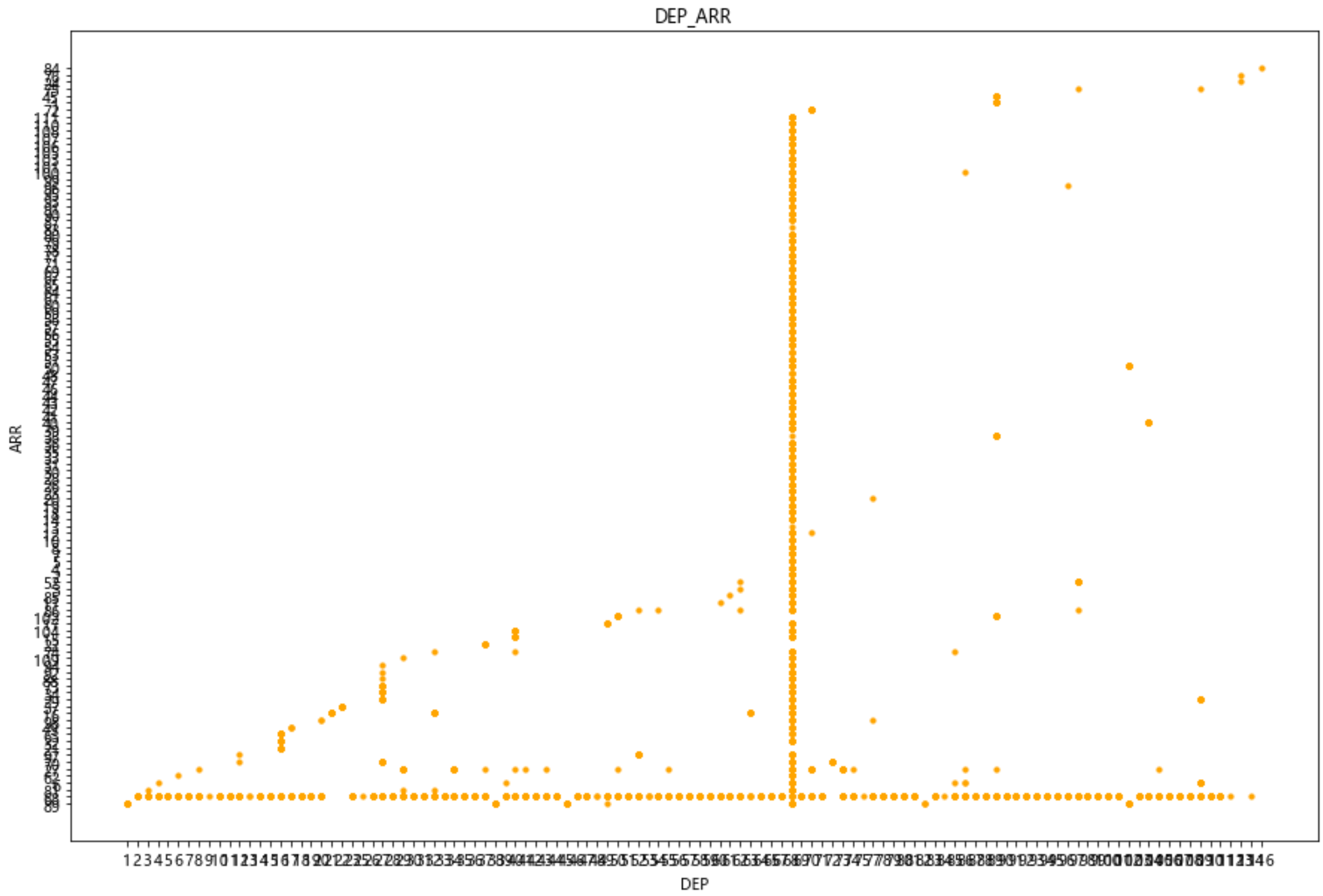

According to the scatter plot, it can be observed that regardless of the value of ARR, there are generally over-limit situations when DEP=68; and when ARR=68, there are generally over-limit situations regardless of the value of DEP.

Therefore, it can be inferred that there may be problems with the infrastructure construction at Airport 68, resulting in multiple occurrences of over-limit situations.

# VI. Model Building and Solving for Problem 4

## 6.1 Problem Analysis

In problem 4, the task is to establish a mathematical model based on Attachment 3, explore a flight technology evaluation method based on flight parameters, and analyze the flight technology of pilots.

To establish a flight technology evaluation method based on flight parameters, machine learning and deep learning techniques can be used to build a classification model that evaluates the pilot's technology based on flight parameters. Possible specific steps are as follows:

(1) Data preprocessing: Perform preprocessing operations such as cleaning, deduplication, and missing value processing on the data to ensure the quality and completeness of the data. If the training effect is not good, data augmentation can also be performed on the dataset.

(2) Feature extraction: Extract features from flight parameters, select representative features, such as average speed, average altitude, heading change rate, etc.

(3) Labeling: Label the pilot's technology level as a label, such as using different numbers or categories to represent different qualification levels.

(4) Model training: Use machine learning algorithms such as decision trees, support vector machines, neural networks, etc. to train the data and build a classification model that can automatically identify the pilot's technology level.
(5) Model evaluation: Use evaluation metrics such as accuracy, recall rate, F1 value, etc. to evaluate the model and verify its classification effect.
During the analysis process, we encountered a question: In the data table of Attachment 3, the first column is "Landing main control", and its value is personnel 1-4 (we believe this is four different pilots); the second column is "Landing main control personnel qualifications", with data being levels, including A, C, F, J, M, T levels (representing Advanced, Competent, Fluent, Intermediate, M-level: Elementary, T-level: Basic).
However, in data processing, we found that the same pilot may correspond to different "qualifications", therefore we believe that this "Landing main control personnel qualifications" refers to **the rating of the operation of the person in a certain flight.**
From the above analysis, we need to establish an evaluation method to rate a pilot's operation based on their flight technology data in a single flight.
When rating, we need to maintain these three principles:
Comprehensive: In data processing, the principle to be maintained is to delete useless indicators, but to retain as much information as possible from the original data, and use a large amount of data and indicators for calculation and evaluation.
Fairness: All personnel's operations are compared under the same evaluation system to prevent rating errors due to inconsistent standards.
Objectivity: Use machine learning and deep learning methods for training data to avoid subjective factors in evaluation.

## 6.2 Data Preprocessing

The data in Attachment 3 is relatively redundant, with many data having missing values, some data being fixed values in the table, and some columns being in an improper format. Therefore, data preprocessing is an essential step before establishing the evaluation system. We use the following rules for data preprocessing:
Delete columns that only have empty values and fixed values (such as aircraft type, TO Gate 1, etc.)
Delete columns that are irrelevant to pilot technology evaluation (such as date)
For columns where most of the data is the same and only a small part of the data is different, if the frequency of occurrence of the small part of the data is less than 10%, delete the column.
For data that varies within a certain range, calculate the variance of the column as part of the evaluation index.

## 6.3 Model Building

We will divide the evaluation system we build into two parts: macro evaluation, which is a long-term overall evaluation of pilots, and micro evaluation, which evaluates the pilot's technology level in each flight process.

### 6.3.1 Macro Evaluation

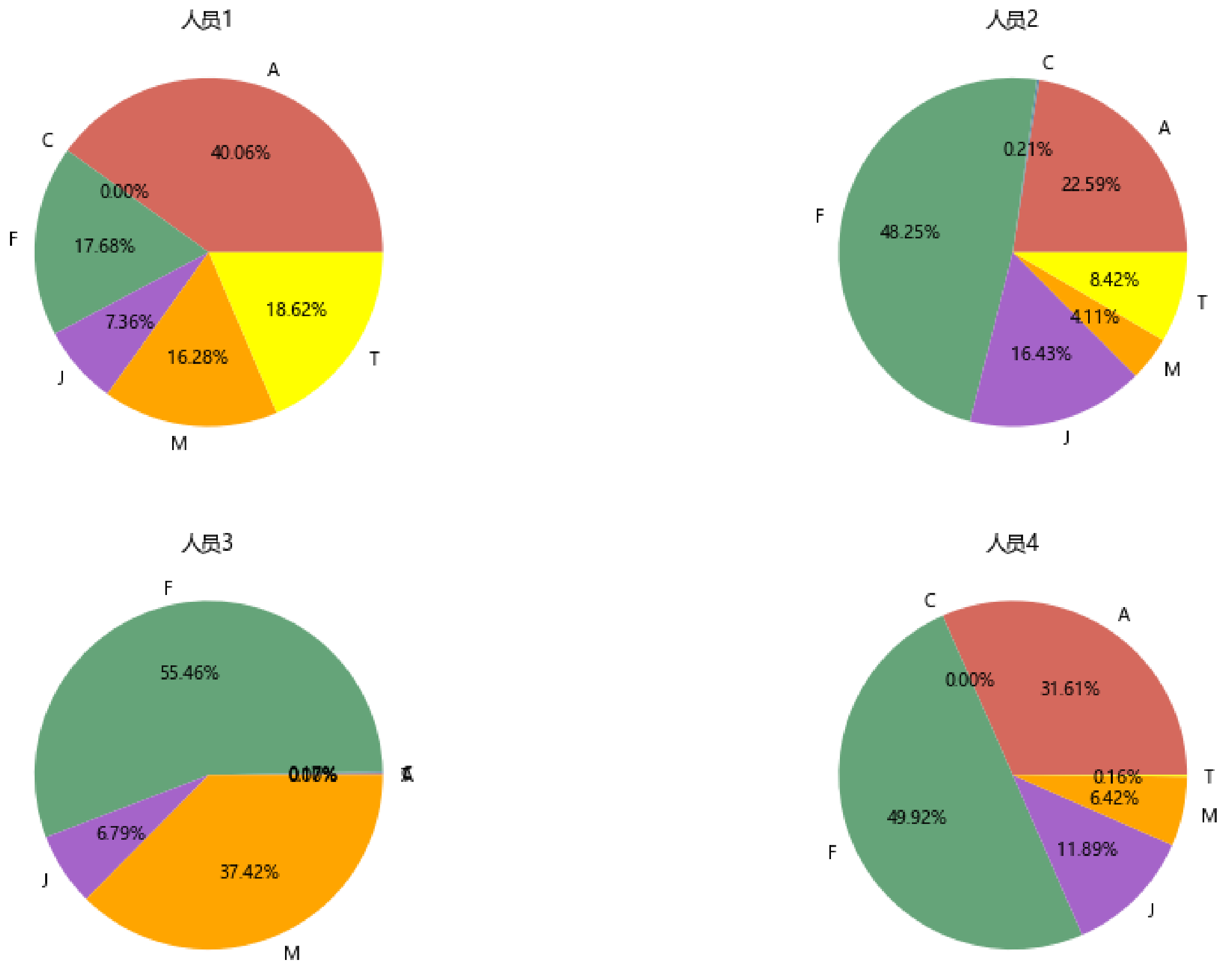

1. Excellent operation
Personnel 1 has the highest A-rating ratio, far higher than personnel 2, 3, and 4. It can be seen that personnel 1 has more excellent operations.
2.Poor operation
Personnel 1 has a T-rating ratio of 18.6%, far higher than personnel 2, 3, and 4, indicating that personnel 1 also has more poor operations.
3.Operation stability
In personnel 3's operations, 55% are F-level operations and 37% are M-level operations. It can be seen that personnel 3 has the best operation stability; personnel 1 has poor operation stability, with a large fluctuation in their operation level rating.

### 6.3.2 Micro Evaluation (Rating)

6.3.2.1 Machine Learning Algorithm Prediction
We first use machine learning to predict the rating of personnel. The algorithms used are: XGBoost algorithm, catBoost algorithm, random forest classification algorithm, GBDT algorithm, decision tree classification, and logistic regression.
The indicators for evaluating the classification prediction results are:
Accuracy: The proportion of correctly predicted samples to the total samples. The higher the accuracy, the better.

Recall: The proportion of positive results actually being positive in the actual positive sample. The higher the recall, the better.

Precision: The proportion of positive results predicted to be positive in the predicted positive sample. The higher the precision, the better.

We will delete the "Landing Master Control" and "Landing Master Control Personnel Qualifications" columns of the dataset as independent variables, and then divide them into training and test sets at a ratio of 8:2. "Landing Master Control Personnel Qualifications" is selected as the dependent variable to be predicted. The evaluation indicators for predicting the effect are accuracy, recall, precision, and F1 value. The results of the operation are as follows.

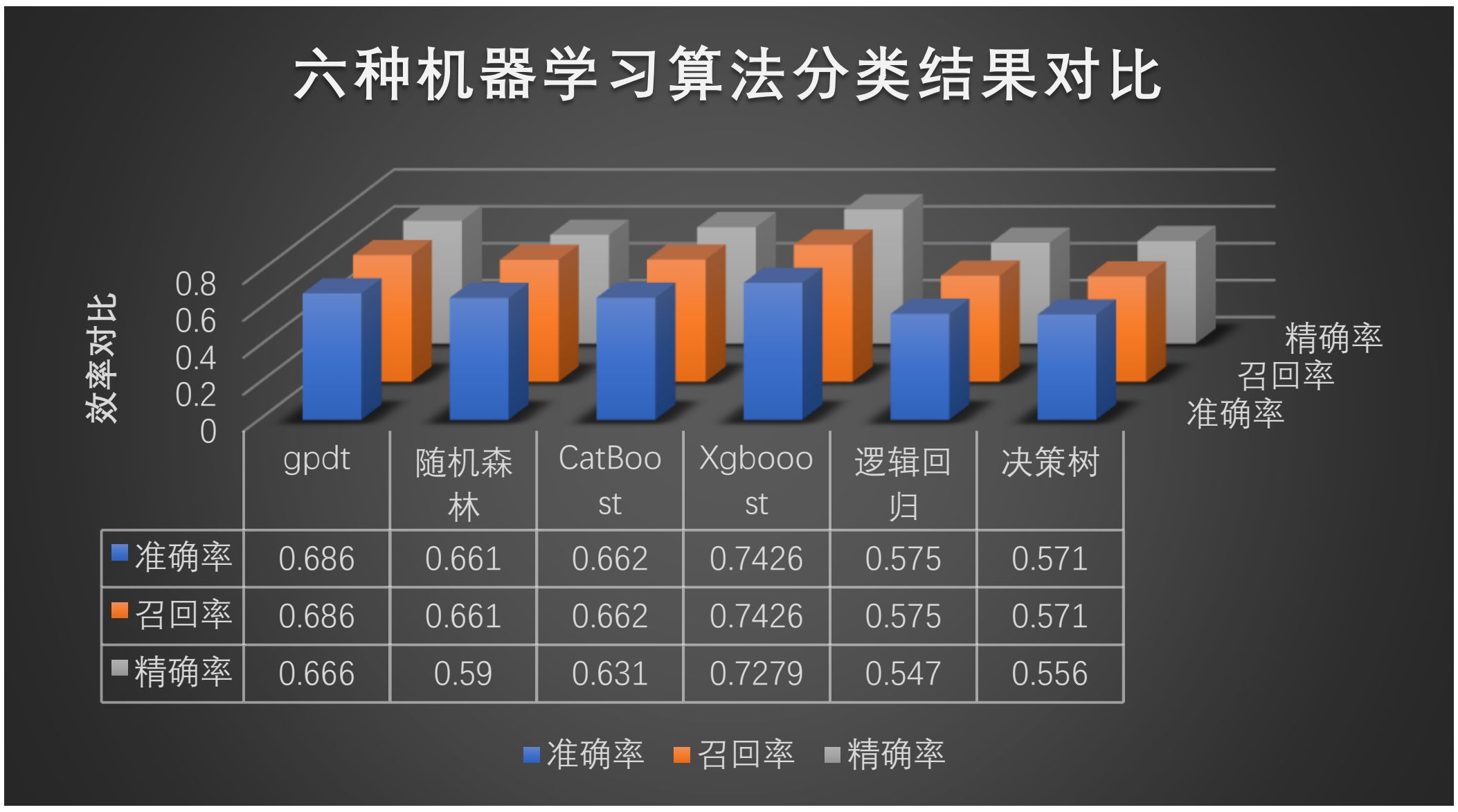

Among them, the algorithm with the highest accuracy rate is the XGBoost algorithm, which reached **74.26%**. The following details the principles and steps of the algorithm.

XGBoost (Extreme Gradient Boosting) is an efficient gradient boosting decision tree algorithm. It improves the original GBDT to enhance the model performance. As a forward additive model, its core is to use the ensemble idea - the Boosting idea, to integrate multiple weak learners into a strong learner through certain methods. That is, multiple trees make decisions together, and the result of each tree is the difference between the target value and the predicted results of all previous trees, and all results are accumulated to obtain the final result, thus achieving the improvement of the overall model performance.

The reason we choose the XGBoost algorithm is that it is composed of multiple CART (Classification And Regression Tree), which can handle classification and regression problems.

The training function process of its tree model is as follows:

The objective function can be described as:

$$Obj = \sum_{i=1}^{n} l(y_i, \hat{y}_i) + \sum_{k=1}^{K} \Omega(f_k)$$

Among them,

$$\sum_{k=1}^{K} \Omega(f_k)$$

Represents the first t-1 trees, which are constant during training, and

$$\sum_{i=1}^{n} l(y_i, \hat{y}_i)$$

Is expanded by Taylor series to define a tree:

$$f_t(x) = w_{q(x)}, \quad w \in \mathbf{R}^T, q : \mathbf{R}^d \rightarrow \{1, 2, \cdots, T\}$$

Then group the leaf nodes to obtain

$$\sum_{j=1}^{T} \left[ \left( \sum_{i \in I_j} g_i \right) w_j + \frac{1}{2} \left( \sum_{i \in I_j} h_i + \lambda \right) w_j^2 \right] + \gamma T$$

Thus, the optimal objective function is obtained:

$$\sum_{j=1}^{T} \left[ G_j w_j + \frac{1}{2} (H_j + \lambda) w_j^2 \right] + \gamma T$$

Using the optimal objective function, the optimal point can be obtained:

$$w_j^* = -\frac{G_j}{H_j + \lambda} \quad \text{Obj} = -\frac{1}{2} \sum_{j=1}^{T} \left[ \frac{G_j^2}{H_j + \lambda} \right] + \gamma T$$

The splitting criterion for leaf nodes is:

$$\text{Gain} = \frac{1}{2} \left[ \frac{G_L^2}{H_L + \lambda} + \frac{G_R^2}{H_R + \lambda} - \frac{(G_L + G_R)^2}{H_L + H_R + \lambda} \right] - \gamma$$

According to the XGBoost algorithm, we use the XGBClassifier() function in sklearn, set the learning rate to 0.05, the number of decision trees to 1000, the maximum depth of trees to 10, specify the loss function as logistic, and finally obtain the training results:

The proportion of feature (independent variable) importance:

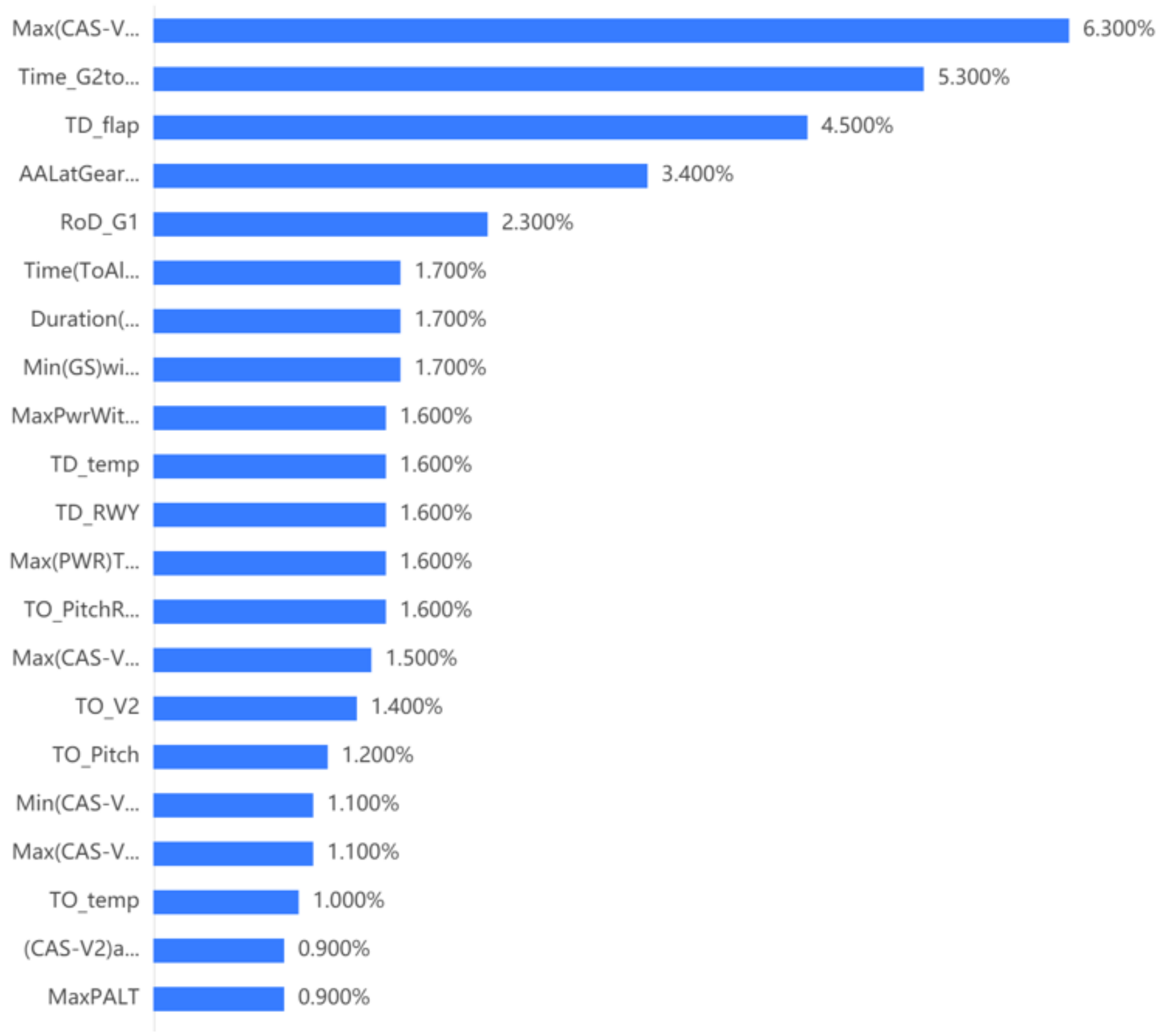

Confusion matrix heatmap:

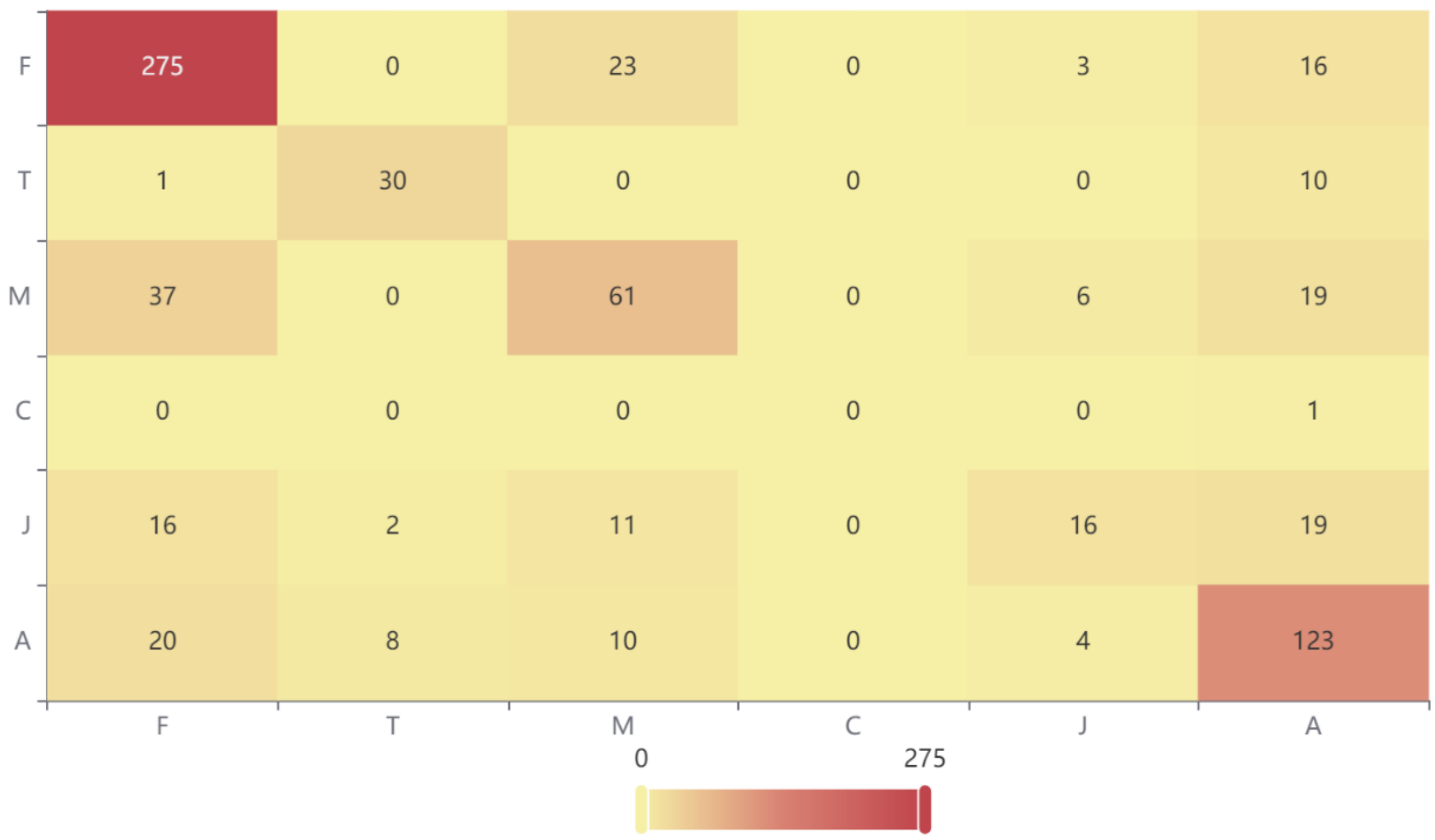

Partial prediction results display:

| 1 | 预测结果Y | 落地主操控 | 预测测试结果概率_A | 预测测试结果概率_C | 预测测试结果概率_F | 预测测试结果概率_J | 预测测试结果概率_M | 预测测试结果概率_T |
|---|---|---|---|---|---|---|---|---|
| 2 | A | A | 0. 59262955 | 0. 003846312 | 0. 32586402 | 0. 064037405 | 0. 008985039 | 0. 004637692 |
| 3 | F | J | 0. 187935 | 0. 003371619 | 0. 44965228 | 0. 041353416 | 0. 30352068 | 0. 014166969 |
| 4 | M | M | 0. 07327815 | 0. 006536602 | 0. 06798813 | 0. 26829147 | 0. 5742528 | 0. 009652905 |
| 5 | F | A | 0. 11399559 | 0. 000707587 | 0. 87500864 | 0. 008412957 | 0. 001209917 | 0. 000665347 |
| 6 | M | M | 0. 14675218 | 0. 009419833 | 0. 12501065 | 0. 31384143 | 0. 39611837 | 0. 008857516 |
| 7 | M | F | 0. 13620026 | 0. 005545031 | 0. 28945956 | 0. 108824305 | 0. 43417197 | 0. 025798881 |
| 8 | F | F | 0. 00344488 | 0. 000481421 | 0. 9534779 | 0. 003715999 | 0. 038286496 | 0. 000593261 |
| 9 | F | F | 0. 062133413 | 0. 001836952 | 0. 9032911 | 0. 02336957 | 0. 007332906 | 0. 002036022 |
| 10 | F | F | 0. 001584666 | 0. 000339425 | 0. 99081415 | 0. 002507347 | 0. 004417562 | 0. 000336822 |
| 11 | F | A | 0. 053398825 | 0. 002540993 | 0. 9098293 | 0. 0205289 | 0. 01013249 | 0. 003569495 |
| 12 | T | T | 0. 030987877 | 0. 001592074 | 0. 00654753 | 0. 015999902 | 0. 008485458 | 0. 93638724 |
| 13 | T | T | 0. 034842283 | 0. 00079045 | 0. 002175905 | 0. 012824919 | 0. 004100293 | 0. 9452662 |
| 14 | A | A | 0. 9516546 | 0. 001977257 | 0. 010314224 | 0. 006514623 | 0. 010680064 | 0. 018859206 |
| 15 | A | A | 0. 70290834 | 0. 003454664 | 0. 22644693 | 0. 04311843 | 0. 007658553 | 0. 016413098 |
| 16 | A | A | 0. 7713801 | 0. 004008099 | 0. 0411961 | 0. 16071518 | 0. 016905084 | 0. 005795401 |
| 17 | A | A | 0. 87490195 | 0. 003559829 | 0. 043459438 | 0. 03103285 | 0. 041499242 | 0. 005546712 |
| 18 | F | F | 0. 01658288 | 0. 000936642 | 0. 96083474 | 0. 005140239 | 0. 012242093 | 0. 004263449 |
| 19 | F | F | 0. 001120569 | 0. 000201711 | 0. 99610186 | 0. 001151104 | 0. 001125761 | 0. 000298997 |

Accuracy metric:

```
预测模型准确率为
Accuracy: 74.26%
```

Recall metric:

```
micro方式召回率为:  0.7426160337552743
macro方式召回率为:  0.6008468973174856
weighted方式召回率为:  0.7426160337552743
```

Precision metric:

```
宏平均精确度为:  0.6985382702403979
微平均精确度为:  0.7426160337552743
加权平均精确度为:  0.7278905984938843
```

The training effect is relatively good.

6.3.2.2 Using Neural Network to Optimize Prediction Algorithm

During the training process of the XGBoost algorithm, we found that the results trained by the algorithm had an accuracy of 1 for indicators such as the training set, but the indicators trained by the testing set were about 0.7. Therefore, we suspect that the overfitting problem was caused by too many training features.

To solve the overfitting problem and optimize the prediction algorithm, we use deep learning to construct a neural network:

We use PyTorch for deep learning and neural network construction. The constructed neural network includes an input layer, a hidden layer, and an output layer. The input layer receives input data, the output layer outputs the network's predicted results, and the hidden layer is responsible for processing and converting the input data to extract useful feature information.

The input layer usually includes multiple input nodes, each node representing a feature of the input data. The hidden layer includes multiple neurons, each neuron receiving a group of input signals, and weighting and adding these signals according to a certain weight and bias. The non-linear transformation is performed by the activation function and finally output to the next layer of neurons or output layer. The output layer includes multiple output nodes, each node representing the network's predicted results for different categories.

The basic structure is shown in the figure below:

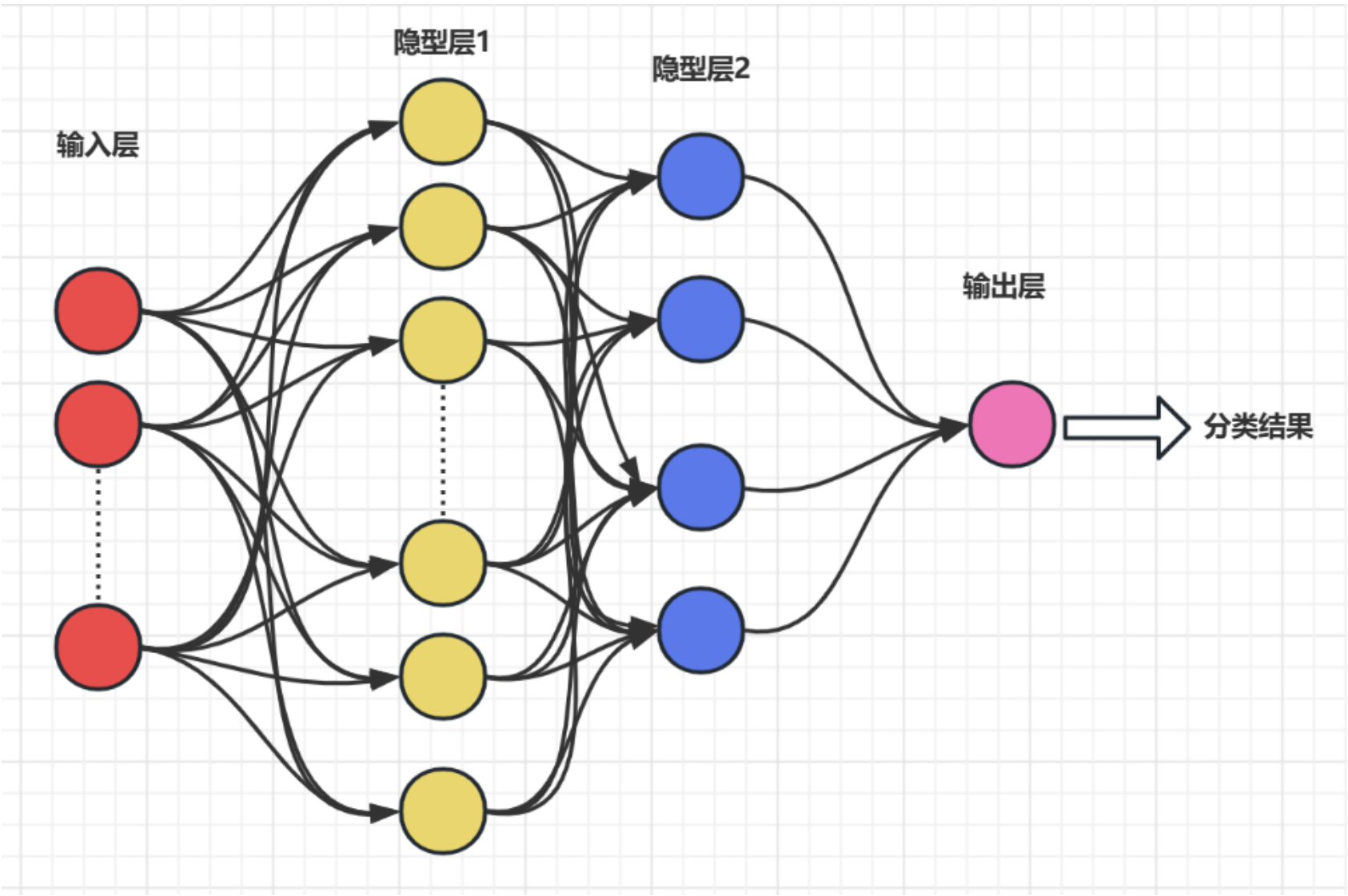

As it is a multi-classification problem, the output layer of the neural network uses the softmax() function to transform the network's output into a probability distribution for classification prediction.

In this neural network, the hidden layer size (hidden_size) is set to 128, and the loss function used is the cross-entropy loss function (CrossEntropyLoss()). The optimizer used is the Adam optimizer. During the training process, stochastic gradient descent (SDG) is used to update the model parameters.

After training, the model is optimized, and the accuracy of the test set prediction obtained after training with the training set is **85.7%**.

```
Accuracy of the network on the test set: 85.70464135021097 %
```

# 7. Model Building and Solution for Question 5

## 7.1 Analysis of Question 5

Question 5 is divided into two parts. Firstly, it requires the establishment of a real-time automated warning system based on the analysis of flight data to prevent safety accidents. Secondly, based on this warning system, a simulation of the data in Attachment 1 is required.

Firstly, according to the names of the data given in Attachment 1, events exceeding the limit that can be calculated based on the Attachment 1 data are selected. As there is no threshold range for the exceeding data in the attachment, this article chooses the monitoring standards for flight quality (QAR) of B737-700 of Air China Limited (2013 Edition) to quantitatively describe each exceeding event. The real-time data is used to calculate the indicators of each exceeding event, which are then compared with the threshold values of the exceeding events to obtain the warning system. Finally, the attachment 1 data is input into the warning system for calculation, and simulation results can be obtained.

## 7.2 Establishment of Automated Warning System

### 7.2.1 Selection of Exceeding Events

7.2.1.1

Some indicators in Attachment 1 can be directly used as measures of exceeding events. The list is as follows:

| 超限事件名称 | 超限阈值 | 附件一中参数 |
|---|---|---|
| 着陆俯仰角大 | ≥8.6 | 姿态（俯仰角） |
| 着陆俯仰角小 | ≤1 | |
| 离地俯仰角大 | ≥10.7 | |
| 下降率大 2000-1000(含)Ft | ≥1500 | 下降率 |
| 下降率大 1000-500(含)Ft | ≥1300 | |
| 下降率大 500-50(含)Ft | ≥1100 | |
| 收起落架晚 | ≥300 | 起落架 |
| 着陆速度 | ≥Vref+15 | 计算空速 |

7.2.2.2

Some of the measurement indicators for exceeding limits cannot be directly obtained from the data in Annex 1, and certain calculations are required to obtain the exceeded events in this section, as shown in the following table:

| 超限事件名称 | 超限阈值 | 附件一中参数 | 计算式 |
|---|---|---|---|
| 爬升速度大 35-1000Ft | ≥V2+30 | 具体时间<br>下降率<br>海拔高度 | (-下降率/60)×3.28 |
| 爬升速度小 35-1000Ft | ≤V2+15 | | |
| 爬升坡度大 35-150Ft | ≥10 | 下降率、地速 | (-下降率/地速)×100 |
| 爬升坡度大 150-400Ft | ≥15 | | |

### 7.2.2 Definition and Explanation of Key Points

Landing: 10 seconds before and after the moment when the air-to-ground switch changes from TRUE to FALSE
Takeoff: 10 seconds before and after the moment when the air-to-ground switch changes from FALSE to TRUE

### 7.2.3 Programming Decision

The landing pitch judgment flowchart is as follows:

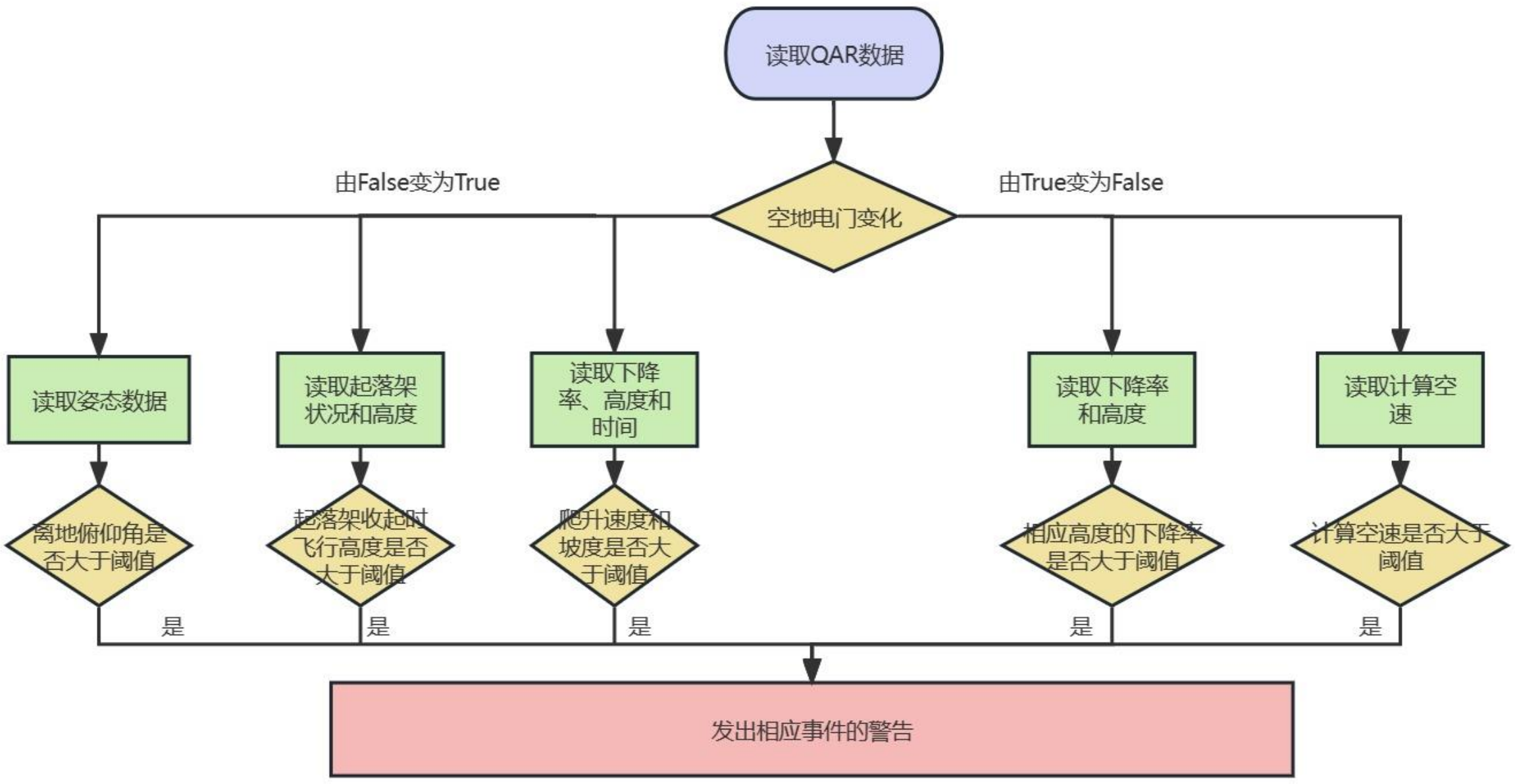

7.3 Simulation Results

| 超限名称 | 事件发生率 |
|---|---|
| 着陆俯仰角大 | 0 |
| 着陆俯仰角小 | 0 |
| 离地俯仰角大 | 0 |
| 着陆速度大 | 12.5% |
| 下降率 2000-1000（含）大 | 0 |
| 下降率 1000-500（含）大 | 0 |
| 下降率 500-50（含）大 | 0 |
| 收起落架晚 | 25% |
| 爬升速度大 | 12.30% |
| 爬升速度小 | 15.87% |
| 爬升坡度大（35-150Ft） | 27.45% |
| 爬升坡度大（150-400Ft） | 18.39% |

# 8 Model Evaluation

## 8.1 Model One

**Advantages:**

No feature scaling required: As it uses a tree structure, different features can be compared without scaling.
Can handle high-dimensional data: Random forests can handle high-dimensional data as they can automatically select the most relevant features, thus avoiding overfitting and improving the model's generalization ability.
Can handle non-linear relationships: Random forests do not assume linear relationships between data, thus can handle non-linear relationships.
**Disadvantages:**
Poor interpretability: Random forests are black-box models, making it difficult to explain their internal decision-making process, which makes them unsuitable for certain fields that require interpretability.
Sensitive to noise: Random forests are sensitive to noise, which can lead to overfitting.

## 8.2 Model Two

**Advantages:**
This model uses the machine learning BP-neural network algorithm to explore only the inherent connections within the data, not influenced by subjective weighting factors, and can maximize the essence of the data.
Small simulation error; this model has excellent fitting effect.
**Disadvantages:**
This model cannot determine the exact mapping relationship between indicators.

## 8.3 Model Four

**Advantages:**
High performance: XGBoost implementation uses efficient data structures and algorithm optimization, can handle large-scale data and high-dimensional features, with high-speed training and prediction.
Interpretability: XGBoost provides feature importance evaluation function, which can help us understand which features have the greatest impact on model performance and select features better.
**Disadvantages:**
Difficult to adjust parameters: XGBoost has many parameters that need to be adjusted, and different parameter combinations may produce different results, making it difficult to judge the optimal parameters.
Sensitive to outliers: XGBoost is more sensitive to outliers, and if there are outliers in the dataset, it may cause model bias.

## 8.4 Model Five

Different countries, airlines, and aircraft models have different QAR monitoring standards in different years, making it difficult to obtain a completely applicable QAR monitoring standard through statistical data, and the generalization ability is poor.
Limited by the data indicators used in Attachment One, the model has insufficient data indicators.